\documentclass[trackchanges,twocolumn]{aastex7}

\usepackage[T1]{fontenc}
\usepackage{threeparttable}
\usepackage[normalem]{ulem}
\usepackage[version=4]{mhchem} 
\usepackage{placeins}
\usepackage{capt-of}
\usepackage{footnote}
\usepackage{siunitx}
\usepackage{textcomp}
\providecommand{\kms}{km~s$^{-1}\,$}
\providecommand{\cms}{cm$^{-2}\,$}
\providecommand{\cmt}{cm$^{-3}\,$}

\usepackage{hyperref}
\begin{document}



\title{An ALMA view on molecular complexity: water, methanol, and methyl formate in the HD~100546 and IRS~48 disks}

\author[orcid=0009-0001-2795-8468]{Luna Rampinelli}
\affiliation{Dipartimento di Fisica, Universit\`a degli Studi di Milano, Via Celoria 16, 20133 Milano, Italy}
\email[show]{luna.rampinelli@unimi.it}
\author[orcid=0000-0003-4689-2684]{Stefano Facchini}
\affiliation{Dipartimento di Fisica, Universit\`a degli Studi di Milano, Via Celoria 16, 20133 Milano, Italy}
\email{stefano.facchini@unimi.it}
\author[orcid=0000-0001-5659-0140]{Marta De Simone}
\affiliation{European Southern Observatory, Karl-Schwarzschild-Straße 2, D-85478, Garching bei München, Germany}
\email{marta.desimone@eso.org}
\author[orcid=0000-0003-3674-7512]{Margot Leemker}
\affiliation{Dipartimento di Fisica, Universit\`a degli Studi di Milano, Via Celoria 16, 20133 Milano, Italy}
\email{margot.leemker@unimi.it}
\author[orcid=0000-0003-2014-2121]{Alice S. Booth}
\altaffiliation{Clay Postdoctoral Fellow}
\affiliation{Center for Astrophysics - Harvard \& Smithsonian, 60 Garden St., Cambridge, MA 02138, USA}
\email{alice.booth@cfa.harvard.edu}
\author[orcid=0000-0002-7695-7605]{Myriam Benisty}
\affiliation{Max-Planck Institute for Astronomy (MPIA), Königstuhl 17, 69117 Heidelberg, Germany}
\email{benisty@mpia.de}
\author[orcid=0000-0003-2045-2154]{Pietro Curone}
\affiliation{Departamento de Astronomía, Universidad de Chile, Camino El Observatorio 1515, Las Condes, Santiago, Chile}
\email{pcurone@das.uchile.cl}
\author[orcid=0000-0001-9549-6421]{Elizabeth, Humphreys}
\affiliation{Joint ALMA Observatory (JAO),Alonso de Córdova 3107, Vitacura 763-0355, Casilla 19001, Santiago, Chile}
\affiliation{European Southern Observatory (ESO),Alonso de Córdova 3107, Vitacura 763-0355, Santiago, Chile}
\email{ehumphre@eso.org}
\author[0000-0003-1859-3070]{Leonardo Testi}
\affiliation{Alma Mater Studiorum Università di Bologna, Dipartimento di Fisica e Astronomia (DIFA), Via Gobetti 93/2, 40129 Bologna, Italy}
\email{leonardo.testi@unibo.it}



\begin{abstract}
Water and methanol are the fundamental constituents of interstellar and cometary ices: understanding their abundances in disks is therefore of high importance to reveal how this ice reservoir evolves across the timeline of star and planet formation. We present ALMA band~5 line emission observations of water and methanol in the warm Herbig disks of HD~100546 and IRS~48. While the IRS~48 disk is methanol-rich, water vapor is not detected, implying a high methanol-to-water column density ratio $>0.2$, possibly reflecting a distinct ice composition. Methanol and water are both detected in the HD~100546 disk, leading to a \ce{CH3OH/H2O} ratio of $\sim0.02$, consistent with the median of the cometary distribution and with the lower edge of the distribution of Young Stellar Objects. We suggest that the gas-phase methanol-to-water ratio in the HD~100546 disk directly reveals the ice reservoir subsequently sublimated in the warm disk stage. Moreover, we report a methyl formate-to-methanol ratio in the IRS~48 disk of $\sim0.2$, consistent with the upper edge of the distribution of Young Stellar Objects. The methyl formate-to-methanol ratio in the HD~100546 disk is $\sim2$, much higher than in younger and older sources, possibly reflecting a chemical enhancement of gas-phase methyl formate. We additionally constrain non-LTE conditions, optical depths, and emitting regions of the methanol lines in both disks. These results offer a new opportunity to probe the chemical trail of molecular complexity in the intermediate disk stage, bridging early and late phases of the star and planet formation history.
\end{abstract}



\section{Introduction}\label{sec:intro}
Water and complex organic molecules (COMs) are basic ingredients for the development of pre-biotically relevant molecules detected in comets and meteorites, and they are primary components of interstellar ices \citep{herbst2009,mumma2011,bockelee-morvan2011}. 
Probing the relative abundance of water and COMs encapsulated within these icy grains is thus fundamental to tracing the evolution of molecular complexity across the star and planet formation history \citep{oberg2021b,ceccarelli2023}, starting from the simplest COM, i.e. methanol. 
A rich statistics on the methanol-to-water ratio has been collected in clouds, Young Stellar Objects (YSOs), and comets from both ice and gas measurements \citep[see e.g.][for a review]{boogert2015,jimenez-serra2025}. While \ce{CH3OH/H2O} ratios are largely consistent among clouds and YSOs, cometary values are slightly lower. This could be indicative of either a variation in the abundance of primordial ice in the solar nebula itself, or of a different degree of ice reprocessing in the protosolar disk \citep{oberg2011ice}. Tracing the intermediate protoplanetary disk stage is thus of primary importance to build a bridge between early and later stages of star formation. In this context, the ice reservoir of water and/or methanol has been suggested to be inherited from early stages of star formation for a few sources \citep[HD~100546, V883~Ori, L1551~IRS5][]{booth2021nat,tobin2023,andreu2023,leemker2025}.

The composition of the ice reservoir can be accessed through infrared absorption spectroscopy \citep[e.g.][]{sturm2024,bergner2026} 
However, this methodology is limited to edge-on disks and is particularly difficult due to e.g., sensitivity limitations and uncertainties on radiative transfer and ice matrices modeling \citep[see e.g.][and refs. therein]{bergner2026}.
 
The sublimation of methanol and water, both characterized by similar desorption temperatures of $\sim150$~K \citep{wakelam2017,minissale2022, Tinacci2023, bariosco2025}, potentially offers an alternative way of revealing the underlying ice-reservoir using gas-phase observations.
In this picture, methanol is known to form only in the ice phase, primarily from hydrogenation of CO-rich ice, while it cannot efficiently form through gas-phase reactions \citep{watanabe2002,garrod2006, geppert2006, fuchs2009, rimola2014, santos2022}. Observations of gas-phase methanol thus reveal the ice content that is released through thermal and/or non thermal processes \citep{walsh2016, booth2021nat, vandermarel2021, temmink2024, evans2025}. This offers a window into the methanol ice budget in protoplanetary disks, an intermediate stage between young protostellar systems and evolved planetary and cometary objects. Combined with observations of sublimated water vapor, these gas-phase methanol observations uniquely allow to reveal if the methanol and water ice content in disks is inherited from early stages of star formation.

However, both methanol and water have relatively high binding energies, which implies that their snowlines are typically close to the central star. In particular, for disks around T~Tauri stars ($M_\star \lesssim 2\, \mathrm{M}_\odot$) this location is within a few au, while it can reach $\sim10$~au for Herbig ($M_\star \sim 2-10\, \mathrm{M}_\odot$) disks \citep{walsh2018, ilee2026}. The molecular complexity budget of the cold outer disk typically probed by sub-millimeter observations is consequently mainly locked into ices, challenging ALMA observations.
Conversely, IR observations can better trace hot gas in the inner regions of protoplanetary disks, but their interpretation is limited by e.g., the presence of multiple spatially unresolved molecular reservoirs and the high optical depth of $\mu$m dust grains hiding a large fraction of the molecular emission \citep[see e.g.][]{temmink2025MINDS, houge2025, leemker2025HLTau}.

Spatially resolved ALMA observations (with mm-dust grains being much more settled to the midplane) can mitigate these issues. In this context, a methanol-to-water ratio of $\sim0.16$ was extracted for the disk of the outbursting V883~Ori source from H$_2^{18}$O and \ce{13CH3OH} \citep{leemker2023}. In this case, the high temperatures $\gtrsim100$~K induce the release of molecules from icy grains out to tens of au, enabling ALMA observations of sublimated water and COMs \citep[see e.g.][and refs. therein]{lee2019, jeong2025, leemker2025, nakasone2026}. Similarly, Herbig disks with wide mm-dust cavities present ideal physical conditions to reveal the COMs and water icy reservoirs sublimated at larger distances from the central star ($\sim10-30$~au), thanks to the intrinsically warmer temperatures and efficient irradiation of the dust cavity \citep{booth2026}. In particular, the methanol reservoir observed in these sources is expected to be directly inherited from early stages of star formation, since the high temperatures of Herbig disks limit the efficient freeze-out of CO onto dust grains, thus preventing in-situ formation of methanol ice \citep{booth2021nat, booth2025HD100453, leemker2023, leemker2024}.
\begin{figure*}[ht!]
    \centering
    \includegraphics[scale=0.23]{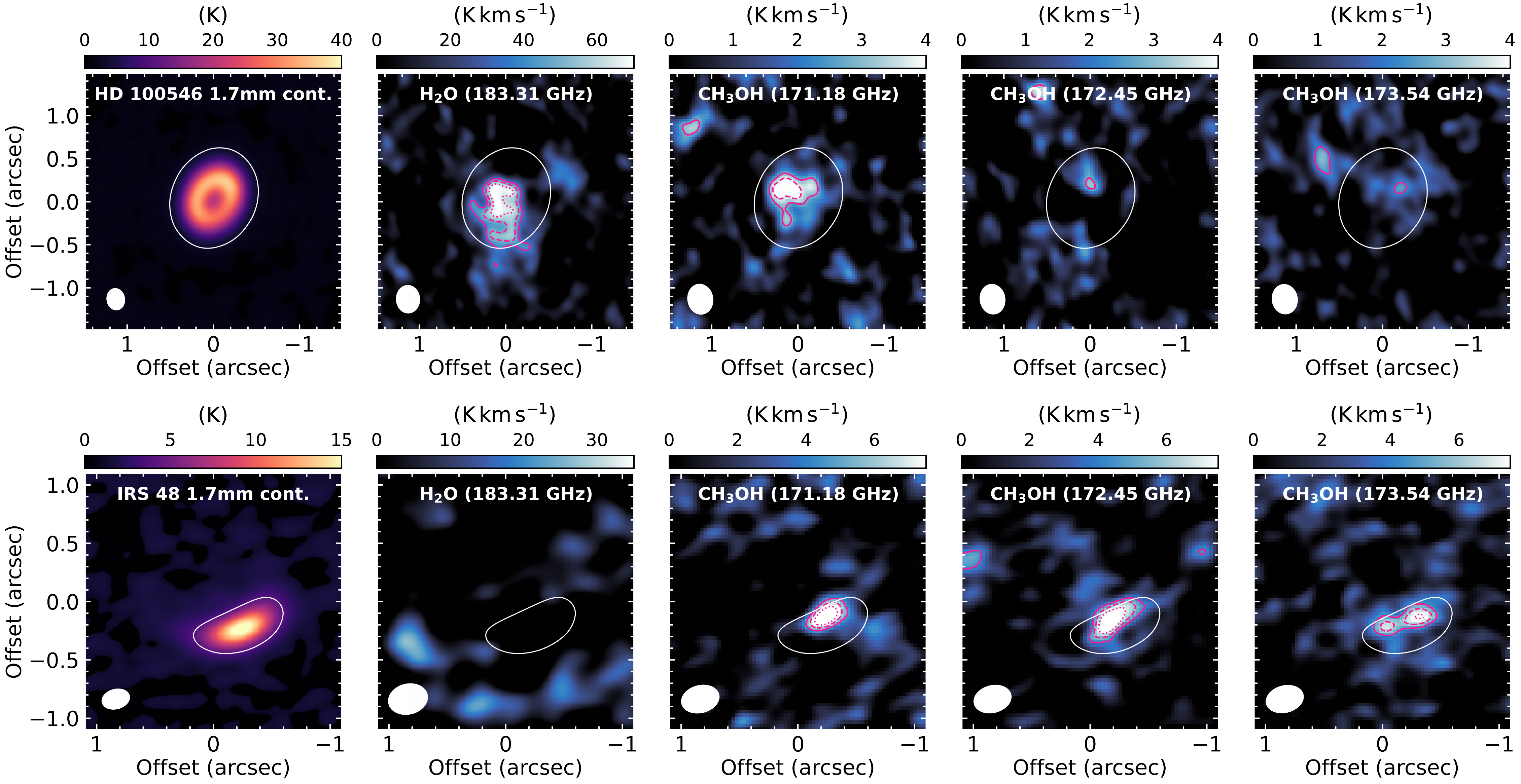}
    \caption{Integrated intensity maps of the 1.7~mm continuum, 183~GHz water line, and three methanol lines in the HD~100546 (top row) and IRS~48 (bottom row) disks. White contours highlight the inner dust ring of HD~100546 and the asymmetric mm-crescent of IRS~48 from their continuum maps, pink contours show 3, 4, and 5$\sigma$ levels. The white ellipse in the bottom left of each panel shows the beam.}
    \label{fig:M0}
\end{figure*}
\cite{rampinelli2026} recently directly constrained the water snowline in the Herbig disk of HD~100546 using spatially-resolved ALMA observations of the main water isotopologue, suggesting that water vapor is primarily produced from thermal sublimation of icy grains at the location of the warm cavity wall. The gas-phase methanol-to-water column density ratio is thus expected to directly trace the underlying ice composition. This opens a new path to address the chemical evolution of molecular complexity in the birthplace of planets, bridging the gap between early and later stages of star and planet formation.

In this work we present water, methanol, and methyl formate ALMA observations in the disks of the Herbigs HD~100546 and IRS~48, comparing the methanol-to-water and methyl formate-to-methanol ratios we extract with early and later stages of star and planet formation.
We describe the sources in Sect.~\ref{sec:source} and the analyzed observations in Sect.~\ref{sec:obs}. We present the analysis on the water vapor, methanol, and methyl formate emission, with the relative ratios, in Sect.~\ref{sec:results}. We discuss the main results in Sect.~\ref{sec:discussion}, and summarize the conclusions in Sect.~\ref{sec:conclusions}.

\section{The sources}\label{sec:source}
\begin{table*}
\centering
  \caption{Listed band~5 imaged lines, corresponding rest frequencies, line parameters (upper state energy, Einstein coefficient, upper state degeneracy), imaging parameters (channel width, beam, $\epsilon$ values from the JvM correction, and RMS), fluxes with corresponding uncertainties (without including the uncertainty on the absolute flux calibration), and critical densities.}
  \label{tab:imaging_molecules}
  \noindent
  \resizebox{\textwidth}{!}{%
  \begin{threeparttable}
          \begin{tabular}{c c c c c c c c c c c}
              \hline
              \hline
              Transition\tnote{a} & Frequency & $E_\mathrm{u}$ & log10(A$_\mathrm{ul}$/s$^{-1}$) & $g_\mathrm{u}$ & $\Delta v$ & Beam & $\epsilon$ & RMS & Flux (integrated intensity)\tnote{b} & $n_\mathrm{cr}$\tnote{c}\\
              & [GHz] & [K] &  & & [km~s$^{-1}$] & bmaj $\times$ bmin (PA) & &[mJy~beam$^{-1}$] & [mJy (beam$^{-1}$)~km~s$^{-1}$] & [\cmt]\\
              \hline
              & & & & HD 100546 & & & & \\
              \hline
              \ce{H2O} ($3_{1,3}-2_{2,0}$) & 183.3101 & 204.71 & -5.445 & 7 & 1.2 & $0\farcs32 \times 0\farcs27 \,(3^\circ)$ & 0.49 & 5 & $490 \pm 122$ & $4\times10^8$\\
              \ce{CH3OH} ($11_{1,10}-11_{0,11}$ E) & 171.1825 & 174.26 & -4.639 & 92 & 1.2 & $0\farcs35 \times 0\farcs29 \,(9^\circ)$& 0.45 & 0.4 & $12 \pm 7$ & $3\times10^6$\tnote{d}\\
              \ce{CH3OH} ($7_{0,7}-6_{1,5}$ E) & 172.4458 & 78.08 & -4.935 & 60 & 1.2 & $0\farcs35 \times 0\farcs29 \,(9^\circ)$ & 0.45 & 0.4 & $< 4.4$\tnote{e} & $7\times10^5$\\
              \ce{CH3OH} ($12_{1,11}-12_{0,12}$ E) & 173.5446 & 202.12 & -4.634 & 100 & 1.2 & $0\farcs35 \times 0\farcs29 \,(8^\circ)$& 0.45 & 0.4 & $17 \pm 7$ & - \\
              \ce{CH3OCHO} ($14_{7,8}-13_{7,7}$ A) & 172.6922 & 94.53 & -4.238 & 58 & 0.9 & $0\farcs35 \times 0\farcs29 \,(9^\circ)$ & & 0.5 & & -  \\
              \ce{CH3OCHO} ($14_{7,8}-13_{7,7}$ A) & 172.6934 & 94.53 & -4.238 & 58 & 0.9 & $0\farcs35 \times 0\farcs29 \,(9^\circ)$ & 0.45 & 0.5 & $32 \pm 8$\tnote{f} & -\\
              \ce{CH3OCHO} ($14_{7,8}-13_{7,7}$ A) & 172.6934 & 94.53 & -4.238 & 58 & 0.9 & $0\farcs35 \times 0\farcs29 \,(9^\circ)$ & & 0.5 & & - \\
              \hline
              & & & & & IRS 48 & & & &  \\
              \hline
              \ce{H2O} ($3_{1,3}-2_{2,0}$) & 183.3101 & 204.71 & -5.445 & 7 & 1.2 & $0\farcs34 \times 0\farcs26 \,(104^\circ)$ & -- & 5 & $<41$\tnote{e} & $4\times10^8$\\
              \ce{CH3OH} ($11_{1,10}-11_{0,11}$ E) & 171.1825 & 174.26 & -4.639 & 92 & 1.2 & $0\farcs34 \times 0\farcs26 \,(104^\circ)$& 0.51 & 0.7 & $13 \pm 1$ & $3\times10^6$\tnote{d} \\
              \ce{CH3OH} ($7_{0,7}-6_{1,5}$ E) & 172.4458 & 78.08 & -4.935 & 60 & 1.2 & $0\farcs34 \times 0\farcs26 \,(104^\circ)$& 0.51 & 0.7 & $11 \pm 2$ & $7\times10^5$\\
              \ce{CH3OH} ($12_{1,11}-12_{0,12}$ E) & 173.5446 & 202.12 & -4.634 & 100 & 1.2 & $0\farcs34 \times 0\farcs26 \,(104^\circ)$& 0.51 & 1.1 & $8 \pm 2$ & -\\
              \ce{CH3OCHO} ($14_{7,8}-13_{7,7}$ A) & 172.6922 & 94.53 & -4.238 & 58 & 1.2 & $0\farcs33 \times 0\farcs23 \,(104^\circ)$ & & 0.7 & & - \\
              \ce{CH3OCHO} ($14_{7,8}-13_{7,7}$ A) & 172.6934 & 94.53 & -4.238 & 58 & 1.2 & $0\farcs33 \times 0\farcs23 \,(104^\circ)$ & -- & 0.7 & $<5.8$\tnote{e,f} & - \\
              \ce{CH3OCHO} ($14_{7,8}-13_{7,7}$ A) & 172.6934 & 94.53 & -4.238 & 58 & 1.2 & $0\farcs33 \times 0\farcs23 \,(104^\circ)$ & &0.7 & & - \\
              \hline
              \hline
          \end{tabular}
          
          \begin{tablenotes}
          \centering
              \item[a] Quantum numbers are formatted as $(J, K_\mathrm{a}, K_\mathrm{c})$ (symmetry) for methanol \citep[CDMS,][]{ CH3OHcmds1}, methyl formate \citep[JPL, ][]{methylformateJPL}, and water \citep[JPL,][]{waterJPL2,waterJPL1}.
              \item[b] Integrated flux (intensity) inside a circle with a radius of 110~au (from the spectrum of the brightest pixel) for HD~100546 (IRS~48).
              \item[c] Critical density calculated for a kinetic temperature of $100$~K. The critical density for the methyl formate lines is not calculated because collisional rates are unavailable.
              \item[d] Critical density calculated for collisions with para-\ce{H2} only, as collisional rates with ortho-\ce{H2} are not available (see Appendix~\ref{appendix:critical_density}).
              \item[e] The upper limit is extracted as $3\times1.1\times \sqrt{\Delta v \, \delta v} \times \mathrm{RMS}(\sigma_\mathrm{s})$ for IRS~48 (HD~100546), where $\Delta v$ is the channel width, $\delta v$ is the line width, and $\sigma_\mathrm{s}$ is the standard deviation across the velocity bins (excluded from the flux calculation) of the integrated spectrum within 110~au. 
              \item[f] The reported flux corresponds to the sum of the blended methyl formate lines.  
          \end{tablenotes}
          \end{threeparttable}
          }
    \end{table*} 

\textit{HD~100546} is a Herbig Ae/Be star ($M_\star =  2.4\, M_\odot$, \citealt{walsh2017}; $L_\star = 23.5\, L_\odot$, \citealt{vioque2018}), at a distance of $108$~pc (\citealt{gaia2023}). 
Sub-millimeter continuum observations and infrared scattered light observations revealed a highly structured dust disk, with two rings at $\sim25$ and $\sim200$~au, and spiral arms \citep{walsh2014, garufi2016, follette2017, sissa2018, pineda2019, fedele2021, maio2025}. Multiple lines of evidence from both sub-millimeter and infrared observations point to the presence of planets in the disk, including also asymmetric molecular emission and kinematic deviations \citep{brittain2014,currie2015,sissa2018,casassus2019,pineda2019,perez2020,booth2023,wolfer2025}. 

\textit{Oph-IRS~48} (hereafter IRS~48) is a Herbig Ae/Be star ($M_\star \sim 2.0\, M_\odot$ and $L_\star \sim 14.3\, L_\odot$, \citealt{brown2012,follette2015,schworer2017,vioque2018}) at a distance of $136$~pc (\citealt{gaia2023}).
Mid-infrared observations show a symmetric disk which becomes more and more asymmetric at longer wavelengths, with ALMA continuum observations revealing an eccentric ring with a prominent dust crescent at a distance of $\sim60$~au from the central star (\citealt{vandermarel2013}; \citeyear{vandermarel2015}; \citealt{ohashi2020}; \citealt{yang2023}). An inner stellar companion has been proposed and discussed as a possible explanation for the observed dust asymmetry \citep{calcino2019, yang2023}.

The disks around both HD~100546 and IRS~48 revealed particularly line-rich spectra with ALMA, with multiple COMs detected, among which are methanol (\ce{CH3OH}) and methyl formate (\ce{CH3OCHO}) (\citealt{vandermarel2021}, \citealt{brunken2022}, \citealt{booth2021nat}, \citeyear{Booth2024HD100546}, \citeyear{booth2024IRS48}, \citealt{leemker2023}, \citeyear{leemker2024}). 
The molecular complexity of these sources combined with spatially resolved ALMA observations of water vapor in the HD~100546 disk \citep{rampinelli2026} thus allows an in-depth characterization of their chemically rich planet-forming environment.

\section{Observations}\label{sec:obs}

In this work, we analyze ALMA observations (12-m only) of HD~100546 and IRS~48 from project 2023.1.01431.S (PI S.~Facchini, band~5) and 2021.1.00738.S (PI: A. S.~Booth, band~7).

The band~7 data were self-calibrated as described by \cite{Booth2024HD100546}. The spectral setup is made of  two spectral settings with four spectral windows (spw) each and with a bandwidth of 1.875 GHz. The spectral windows are centered at 338.790824, 340.732413, 348.916936, and 350.775389 GHz for one setting and 344.240980, 345.940999, 354.367095, and 356.067114 GHz for the second setting.

The spectral setup of the band~5 data is made of two spectral windows (spw) centered on the water $3_{1,3}-2_{2,0}$ line at $\nu = 183.310$~GHz and the \ce{H^13CO^+} $2-1$ line at $\nu = 173.507$~GHz, respectively with a 117~MHz bandwidth each, and two continuum spws centered at 172.0 and 185.5~GHz with a 1.875~GHz bandwidth each. 
We self-calibrated the band~5 data after the ALMA Pipeline cross-calibration as described by \cite{rampinelli2026}, following the Pipeline designed for the exoALMA Large Program \citep{loomis2025}, and using the \texttt{CASA} software version 6.5.4 \citep{CASA2022}. 
We used elliptical masks to place the cleaned component during self-calibration, with Position Angle $\mathrm{PA}=100.0^\circ$, a major axis of $1\farcs5$, and a minor axis of $1\farcs5 \times \cos{(i)}$, where $i = 50.0^\circ$ is the disk inclination, for IRS~48 \citep{bruderer2014}, and $\mathrm{PA}=146.0^\circ$, a major axis of $2\farcs5$, a minor axis of $2\farcs5 \times \cos{(i)}$, where $i = 41.7^\circ$, for HD~100546 \citep{walsh2017}.
Within these masks we calculate a band~5 continuum flux density of 28 and 286~mJy, for the disks of IRS~48 and HD~100546, respectively \citep[consistently with what has been already reported by][for HD~100546]{rampinelli2026}. We report an RMS of 20.4 and 19.3~$\mu$Jy~beam$^{-1}$, and a peak S/N ratio of 614 and 2464, for the disks of IRS~48 and HD~100546, respectively, at the end of the self-calibration. 
The corresponding beams are $0\farcs25 \times 0\farcs17 \,(-76.3^\circ)$ and $0\farcs26\times 0\farcs21 \,(9.8^\circ)$.

We imaged the self-calibrated ms tables with the \texttt{CASA} software \citep{CASA2022} using the \texttt{tclean} task \citep{hogbom1974aperture, cornwell2008} to reconstruct the spectral cubes analyzed in this work. We used the \texttt{multiscale} deconvolver with scales [0,5,10,20,30] pixels, a pixel size 
of $0\farcs01$, and an intensity threshold of $3\sigma$. To place the CLEAN components we used the same elliptical masks built during the self-calibration. Similarly to what was presented in \cite{rampinelli2026}, we imaged the lines with \texttt{Briggs} weighting spanning a range of robust parameters between $0.5$ and $2.0$, and with a channel width of 0.9 and 1.2~\kms. We accounted for the so-called JvM effect \citep{jorsater1995} by applying a correction as described by \cite{czekala2021}, to correctly recover the line fluxes. The cubes used in this work were not primary beam corrected, since the spatial extent of the emission is significantly more compact than the primary beam for both disks. We used cubes imaged with natural weighting in the analysis, and applied an additional $uv$-taper of $0\farcs04\times0\farcs085\, (-75^\circ)$ to the methanol cubes of the IRS~48 disk to achieve the same spatial resolution of the band~7 data (see Sect.\ref{subsec:methanol}). Image properties are listed in Table~\ref{tab:imaging_molecules}.

Figure~\ref{fig:M0} shows the intensity map of the band~5 continuum and the integrated intensity (moment~0) maps of the water line and three methanol lines in HD~100546 (top row) and IRS~48 (bottom row), obtained using the \texttt{bettermoments} package \citep{teague2019bettermoments}.
Corresponding line parameters, imaging specifics, and line fluxes are listed in Table~\ref{tab:imaging_molecules}.

\section{Results}\label{sec:results}

\begin{figure*}[ht!]
    \centering
    \includegraphics[scale=0.3]{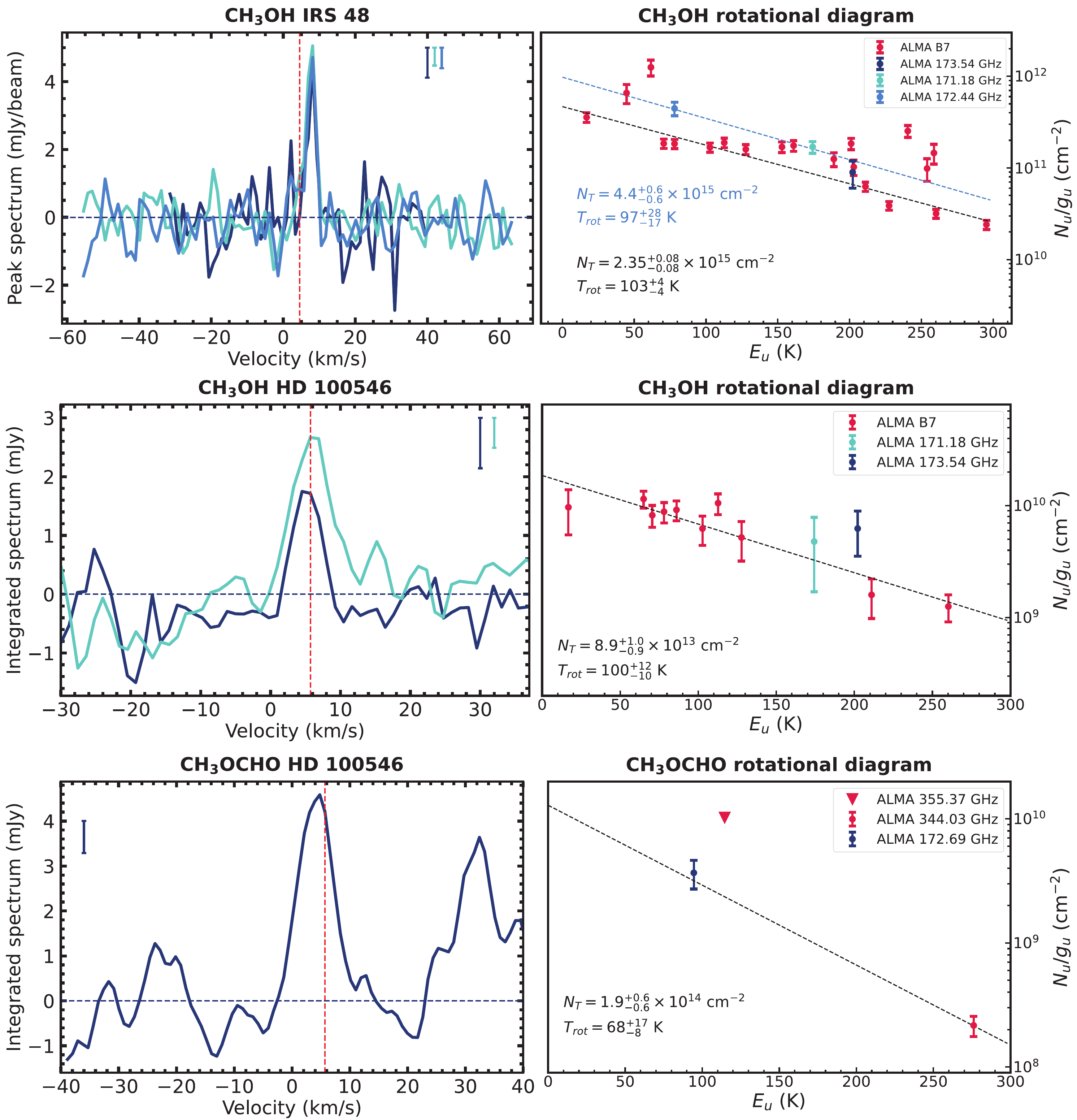}
    \caption{Left: peak and disk-integrated spectra of the band~5 methanol and methyl formate lines detected in the IRS~48 and HD~100546 disks (from top to bottom), with blue bars indicating the uncertainty for each line spectrum. Right: rotational diagrams of the methanol and methyl formate (from top to bottom). Best fit results are indicated by the dashed lines. For the case of IRS~48 we compare the results using both band~7 and 5 methanol lines (black) and only band~5 transitions (light blue).}
    \label{fig:spectrum_rotdiag_CH3OH}
\end{figure*}

\subsection{Non-detection of water vapor in IRS~48}\label{subsec:water_IRS48}
While the water line at 183.31~GHz is detected in the HD~100546 disk \citep[][and see the second panel in the top row of Fig.~\ref{fig:M0}]{rampinelli2026}, the same line is not detected in the IRS~48 disk, with an RMS of 5 mJy beam$^{-1}$ corresponding to the image properties listed in table~\ref{tab:imaging_molecules}. The second panel in the bottom row of Fig.~\ref{fig:M0} shows the integrated intensity map of the water line obtained from the cube imaged with natural weighting and a channel width of 1.2~\kms. The white contours indicate the location of the dust crescent (see the first panel in the bottom row of Fig.~\ref{fig:M0}). Figure~\ref{fig:channel_maps} in Appendix~\ref{appendix:water_IRS48} shows an example of the channel maps of the undetected water line. We additionally applied techniques to boost the effective signal-to-noise of the data in the image- and $uv$-plane, which did not lead to a detection of the water line (see Appendix~\ref{appendix:water_IRS48} for more details).

We extracted the upper limit on the water flux (see Table~\ref{tab:imaging_molecules}) from the RMS as $3\times1.1\times \sqrt{\Delta v \, \delta v} \times \mathrm{RMS}$, where $\Delta v$ is the channel width ($1.2$~\kms) and $\delta v$ is the line width (assumed to be 6~\kms as informed by the methanol lines).

\subsection{Methanol}\label{subsec:methanol}
Methanol lines were previously detected in ALMA band~7 observations of the two disks \citep{booth2021nat,vandermarel2021,brunken2022,booth2024IRS48,Booth2024HD100546,temmink2024,evans2025} and three additional methanol lines at 171.18, 172.45, and 173.54~GHz are covered by the band~5 setup (see Table~\ref{tab:imaging_molecules}).
In particular, in the disk of HD~100546 we detect the line at 171.18~GHz with a peak signal-to-noise (S/N) in the spectrum of $5.2$ and the 173.54~GHz line with a $\mathrm{S/N} = 3.3$. The spectra of the two lines corresponding to the reported S/N ratios are shown in the middle-left panel of Fig.~\ref{fig:spectrum_rotdiag_CH3OH}. The spectra are extracted from a projected circle with a radius of 55~au and correcting for the Keplerian rotation of the disk \citep{teague2019gofish}. This boosts the S/N ratio highlighting the detections, including the faint line at 173.54~GHz, that would otherwise not be visible from the integrated intensity map in Fig.~\ref{fig:M0}. A similar approach to confirm the detection of the lines that are consistent with the Keplerian rotation of the disk is described in Appendix~\ref{appendix:teardrop_methanol} and shown in Fig.~\ref{fig:teardrop_methanol}.

For the IRS~48 disk we detect all the three methanol lines at 171.18~GHz, 172.45~GHz, and 173.54~GHz with a peak S/N ratio of 9.5, 7.7, and 5.1, respectively.
The spectra corresponding to the reported S/N ratios are shown in the top-left panel of Fig.~\ref{fig:spectrum_rotdiag_CH3OH} and are extracted from the brightest pixel without correcting for Keplerian rotation, since the emission is asymmetric and localized at the mm-crescent. The spectrum is shifted with respect to the systemic velocity as it corresponds to a de-centered pixel coincident with the mm-crescent.

The line fluxes (integrated intensity) with the relative uncertainties are reported in Table~\ref{tab:imaging_molecules} and are calculated within a circle with a radius of 110~au as in \citealt{evans2025} (from the location of the brightest pixel - same for each line - as in \citealt{temmink2024}) for HD~100546 (IRS~48). 
We highlight that the significance of the total fluxes for the HD~100546 disk is lower than $3\sigma$, because we used a large integration area, to be consistent with \cite{evans2025}. The fluxes extracted within a smaller area with a radius of 55~au have a higher significance $>3\sigma$. The fluxes are obtained by integrating the spectra over a velocity range of 4 to 10~\kms for IRS~48 and 0 to 10~\kms for HD 100546, after visual inspection of the data. The uncertainty on the line fluxes is obtained as $\sigma_\mathrm{s} \times \Delta v \times \sqrt{N_\mathrm{bins}}$, where $\sigma_\mathrm{s}$ is the standard deviation across the velocity bins not included in the integration, $\Delta v$ is the width of the velocity bins, and $N_\mathrm{bins}$ are the number of bins integrated to extract the flux. We used the spectrum extracted from the JvM-corrected cube to calculate the line flux, and the non-JvM-corrected one for the related uncertainty, following \cite{rampinelli2025}, not to underestimate the uncertainty \citep{czekala2021, casassus&carcamo2022}.

From the line fluxes obtained from ALMA band~7 \citep[see ][ and Table~\ref{tab:b7_methanol}]{temmink2024, evans2025} and band~5 (see Table~\ref{tab:imaging_molecules} in Appendix~\ref{appendix:b7_methanol}) observations, we extracted the methanol molecular column density $N_\mathrm{t}$ and excitation temperature $T_\mathrm{ex}$ in the two disks. We included a 10\% absolute flux calibration uncertainty in the calculation. We thus applied the rotational diagram (RD) analysis \citep{goldsmith1999population}, as described in, e.g., \cite{rampinelli2024, evans2025}, under the assumption of optically thin emission and of a homogeneous excitation temperatures among the lines and within the emitting area. These assumptions are discussed by \cite{temmink2024} and \cite{evans2025}, and will be addressed in more details in Sect.~\ref{sec:discussion}. For the disk of HD~100546 we assumed a circle with a radius of 110~au as emitting area, consistent with \cite{evans2025}, while we assumed that the emitting area is equal to the beam area for the case of IRS~48 \citep{temmink2024}. The results of the rotational diagram analyses are presented in the top- and middle-right panels of Fig.~\ref{fig:spectrum_rotdiag_CH3OH}.

For the disk of HD~100546, we obtain $N_\mathrm{t} = 8.9^{+1.0}_{-0.9} \times 10^{13}$~\cms and $T_\mathrm{ex} = 100^{+12}_{-10}$~K, that is consistent with the result presented by \cite{evans2025} of $N_\mathrm{t} = 1.4^{+0.4}_{-0.3} \times 10^{14}$~\cms and $T_\mathrm{ex} = 152^{+35}_{-27}$~K, within $1.5\sigma$, where they used only methanol lines detected in ALMA band~7 observations. We note that the small difference between the results is explained by a difference in the Einstein coefficients of two methanol lines listed by \cite{evans2025} in their Table~1, and the values we took from the CDMS catalogue\footnote{List of molecules and corresponding line parameters from CDMS: https://cdms.astro.uni-koeln.de/classic/entries/}. \cite{evans2025} found a warm ($T_\mathrm{ex}\sim 152$~K) thermally desorbed component within $\sim110$~au. We highlight that the spatial resolution of the band~7 observations does not allow to resolve the morphology of the methanol emission and thus establish whether the peak is spatially coincident with the one of the water vapor. Moreover, they found a cold ($T_\mathrm{ex}\sim 52$~K) photodesorbed component emitting between $\sim180$ and $260$~au. However, we did not apply the RD analysis between $\sim180$ and $260$~au as the two band~5 methanol lines are not detected in this region. This is consistent with the detection of only cold band~7 methanol lines $E_\mathrm{u} < 60$~K in this outer region \citep{evans2025}.

For the disk of IRS~48, we obtain $N_\mathrm{t} = (2.35\pm0.08) \times 10^{15}$~\cms and $T_\mathrm{ex} = 103^{+4}_{-4}$~K. The column density is an order of magnitude higher than what is reported by \cite{temmink2024}. 
Our result for IRS~48 is consistent with what was previously obtained by \cite{booth2024IRS48} and \cite{vandermarel2021} with the same methodology we applied in this work, and with an independent LTE model of the methanol lines with the CASSIS code \citep{brunken2022,leemker2023}.

Because of the large spread of the datapoints in the RD analysis of IRS~48 (see the top-right panel of Fig.~\ref{fig:spectrum_rotdiag_CH3OH}), we additionally applied the RD analysis to the band~5 methanol lines alone, obtaining $N_\mathrm{t} = (4.4\pm0.6) \times 10^{15}$~\cms and $T_\mathrm{ex} = 97^{+28}_{-17}$~K (see dashed blue line in the top-right panel of Fig.~\ref{fig:spectrum_rotdiag_CH3OH}). While the excitation temperature is consistent with the rotational diagram obtained combining band~5 and 7 lines, the column density is a factor of $\sim2$ larger.

\subsection{Non-LTE LVG analysis of methanol}\label{subsec:nonLTE}
\begin{figure*}
    \centering
    \includegraphics[width=0.8\linewidth]{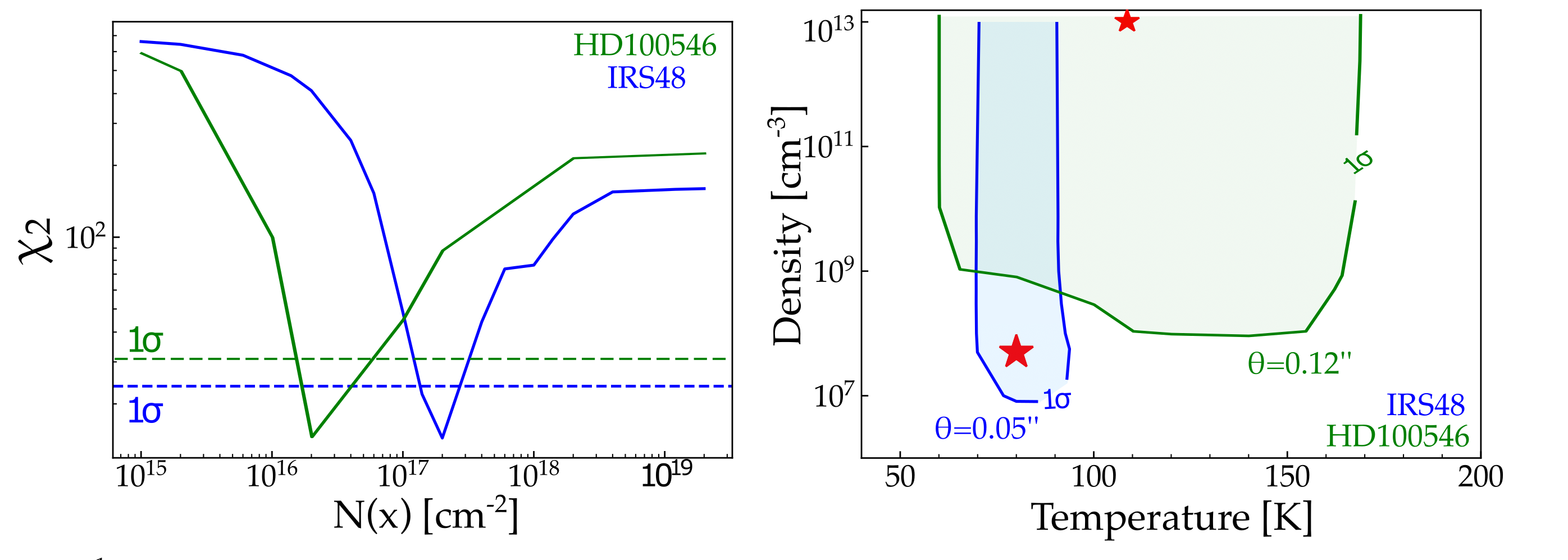}
    \caption{Non-LTE LVG fitting results for methanol (CH$_3$OH) for HD 100546 (in green) and IRS48 (in blue) using lines from this work and from \citet{evans2025} and \citet{temmink2024}. Left: $\chi^2$ minimization for the total column density $N$(CH$_3$OH) (A-type plus E-type). Right: density–temperature $\chi^2$ contour plots showing 1$\sigma$ confidence level, assuming the best-fit total column density of 2$\times10^{16}$ and 2$\times 10^{17}$ cm$^{-2}$,  and best-fit source size of $0\farcs12$ and $0\farcs05$, for HD 100546 and IRS 48 respectively.  The best-fit solutions are marked by red stars.}
    \label{fig:LVG_results}
\end{figure*}

\begin{table*}[]
    \centering
    \caption{Best-fit results and 1$\sigma$ Confidence Level (range) from the non-LTE LVG analysis of CH$_3$OH in IRS 48 and HD 100546.}
    \label{tab:LVG_results} 
    \begin{tabular}{c|cccccccc}
    \hline
    \hline
         & $n$(H$_2$) [cm$^{-3}$] & $T_\mathrm{kin}$ [K] & $N$(CH$_3$OH) [cm$^{-2}$] & emitting size $\theta$ [$''$] \\
         \hline
         IRS~48  & 5$\times10^7$ ($\geq$ 1$\times10^7$) & 80 (70 -- 95) & $2\times 10^{17}$ ($1.5-3 \times 10^{17}$) & $0\farcs05$ ($0\farcs04 - 0\farcs06$) \\
         HD~100546 & 1$\times10^{13}$ ($\geq$ 1$\times10^8$) & 110 (65 -- 170) & $2\times 10^{16}$ ($1.5 -7 \times 10^{16}$) & $0\farcs12$ ($0\farcs9 - 0\farcs15$) \\
         \hline
    \end{tabular}
\end{table*}

While the RD analysis provides a first estimate of the physical conditions under the assumption of an homogeneous rotational temperature and optically thin emission, it does not account for possible optical depth effects, nor for the emitting size, which is unknown given that the emission seems to be unresolved in the radial direction within the beam. 
Methanol is one of the few relatively complex species that is well characterized in terms of collisional coefficients (available for transitions with $E_{\rm up} \lesssim 200$ K). This allows us to
perform a more accurate non-LTE analysis using a Large Velocity Gradient (LVG) code \citep[\texttt{grelvg};][]{ceccarelli2003}, which properly accounts for optical depth effects and includes the treatment of the emitting size.

We used the collisional coefficients of both symmetry species of CH$_3$OH \citep[A and E;][]{rabli_rotational_2010} with para-H$_2$, computed between 10 and 200 K for the first 256 levels and
provided by the BASECOL database \citep{dubernet_basecol2012:_2013}. We assumed a spherical geometry to compute the line escape probability \citep{de_jong_hydrostatic_1980}, a CH$_3$OH-A/CH$_3$OH-E ratio of 1, an H$_2$ ortho-to-para ratio of 3. 

We ran a large grid of models ($>10^4$) covering the observed frequencies, for HD~100546 (IRS~48) varying the total \ce{CH3OH-A}~+~\ce{CH3OH-E} column density $N({\rm CH}_3{\rm OH})$ from $2\times10^{12}$ to $2\times10^{19}$ cm$^{-2}$  ($1\times10^{15}$--$2\times10^{19}$ cm$^{-2}$), the gas density $n_{\rm H_2}$ from $5\times10^{7}$ to $5\times10^{13}$ cm$^{-3}$ ($10^{6}$--$10^{10}$ cm$^{-3}$), and the temperature $T$ from 30 to 200 K (40--200 K). We simultaneously fitted the observed CH$_3$OH-A and CH$_3$OH-E line intensities by comparing them with the LVG predictions, leaving $N({\rm CH}_3{\rm OH})$, $n_{\rm H_2}$, $T$, and the emitting size $\theta$ as free parameters. As informed by observations, we assumed FWHMs of 4 km s$^{-1}$ and 3.5 km s$^{-1}$ for HD~100546 and IRS~48, respectively, and included a 15\% calibration uncertainty on the observed integrated intensities. The results are insensitive to variations in the linewidth of $\pm 0.5$ km s$^{-1}$.

For the IRS~48 disk, the best fit is obtained for a size (diameter) of
$\theta = 0.05''$, $N({\rm CH}_3{\rm OH}) = 2\times10^{17}$ cm$^{-2}$, $T_{\rm gas} = 80$ K, and $n_{\rm H_2} > 10^{7}$ cm$^{-3}$ (i.e. levels close to LTE), with a reduced $\chi^2 \sim 1.0$. Five lines are found to be optically thick, with optical depths up to $\tau \sim
4.7$. We highlight that \cite{temmink2024} also suggested a similarly small emitting region with size $\lesssim8$~au for the IRS~48 disk, in order for the methanol lines to be optically thick.

For HD~100546, the best fit (reduced $\chi^2 = 0.84$) is obtained for $N({\rm CH}_3{\rm OH}) = 2\times10^{16}$ cm$^{-2}$, $\theta = 0.12''$, $T_{\rm gas} = 110$ K, and $n_{\rm H_2} > 10^{8}$ cm$^{-3}$ (again consistent with LTE populations), with all lines optically thin ($\tau \sim 0.07-0.24$). Critical densities of the methanol lines used in this analysis are listed in Table~\ref{tab:imaging_molecules} and \ref{tab:b7_methanol}, and were calculated as described in Appendix~\ref{appendix:critical_density}. These results confirm a posteriori that LTE is a good approximation, while still benefiting from the more robust treatment of radiative transfer and opacity provided by the non-LTE LVG analysis.
An additional advantage of the LVG analysis is that it provides an estimate of the emitting size, which is otherwise unconstrained. Using this information, we can revisit the rotational diagram results by correcting for the dilution factor (ff $= \theta_\mathrm{em}^2$ / ($\theta_\mathrm{em}^2 + \theta_\mathrm{int}^2$), where $\theta_\mathrm{int}$ and $\theta_\mathrm{em}$ are the sizes of the integration and emitting area, respectively). After applying this correction, we find that for HD~100546 the RD results are consistent with the LVG results, confirming that the LTE and optically thin assumptions are valid in this case. In contrast, for IRS~48 the column density inferred from the RD analysis remains lower by a factor of $\sim 4$, reflecting the impact of moderate optical depths ($\tau < 5$) affecting several transitions. This comparison highlights that, while the rotational diagram approach provides a reasonable first-order estimate, it can significantly underestimate the column density when optical depth effects are non-negligible.

Note that the values reported above for  HD~100546 are derived excluding the line at 350.9051 GHz ($E_{\rm up} = 17$ K). This line prevents the model from finding a good solution and leads to unrealistically large emitting sizes.
When constraining the emitting size to be smaller than or equal to the beam, this transition appears as a clear outlier. This is not surprising, as its low excitation energy implies that it may trace multiple gas components with different physical conditions, which cannot be easily disentangled spatially or spectrally, given the spatial resolution of the observations and the large integration area (diameter of $2''$, see Sect.~\ref{subsec:methanol}). We therefore decided to exclude it from the fitting.
We also highlight that the spread in the derived column density and temperature is relatively large. This is likely due to the large integration area. Given that the emission is only marginally resolved (beam $\sim 0.3''$), the spatial dilution is significant, and some level of contamination cannot be excluded.
However, we adopted this integration area to allow a direct comparison with the data and results presented by \cite{evans2025}. We thus expect that a smaller integration region would likely yield tighter constraints, but without significantly affecting the best-fit values. 

Figure \ref{fig:LVG_results} shows, for both sources, the gas density--temperature ($n_{\rm H_2}$--$T$) 1$\sigma$ confidence level at the best-fit $N({\rm CH}_3{\rm OH})$ and best fit emitting size. 
The fit results together with the 1$\sigma$ confidence intervals are summarized in Table \ref{tab:LVG_results}.

\subsection{Methyl formate}\label{subsec:methyl_formate}
We report the detection of three spectrally blended methyl formate lines at 172.69~GHz (see Table~\ref{tab:imaging_molecules} and bottom-left panel of Fig.~\ref{fig:spectrum_rotdiag_CH3OH}) in the HD~100546 disk, with a total flux of ($32\pm8$)~mJy~\kms, within a circle with a radius of 110~au. We additionally report a marginal detection ($\mathrm{S/N}=3$) of a \ce{CH3OCHO} line at 355.37~GHz.
No \ce{CH3OCHO} lines are detected in the IRS~48 disk in band~5 and we report an upper limit on the total flux of the 172.69~GHz line of 5.8~mJy~\kms (see Table~\ref{tab:imaging_molecules}).

Similarly to what we did for methanol, we applied the RD analysis to the methyl formate lines detected in band~5 (excluding the marginally detected line at 355.37~GHz) and 7 in the disk of HD~100546 (\citealt{Booth2024HD100546}), under the assumption of optically thin emission. In particular, we measure a total line flux for the band~7 blended lines at 344.03~GHz of $(63\pm10)$~mJy~\kms, within the same area adopted for the band~5 lines (see Table~\ref{tab:b7_methanol}). Since the line parameters of the blended transitions both in band~7 and 5 are nearly identical (see Table~\ref{tab:imaging_molecules} and \ref{tab:b7_methanol}), we divided the total flux by the number of blended lines (4 for the band~7 and 3 for the band~5 ones). We only considered the brightest lines in the calculation, while we neglected the contribution of fainter blended lines for which the intensity ratio with respect to the brightest transition is expected to be $<15\%$ for band~7 and $<2\%$ for band~5. The result of the rotational diagram analysis is shown in the bottom-right panel of Fig.~\ref{fig:spectrum_rotdiag_CH3OH}: we find $N_\mathrm{t} = (1.9\pm 0.6)\times10^{14}$~\cms and $T_\mathrm{ex}=68^{+17}_{-8}$~K. The marginally detected line at 355.37~GH is displayed as an upper limit.

Methyl formate was previously detected in only a few disks with ALMA, including the outbursting V883~Ori and V1057 cyg sources \citep{lee2019,yamato2024v883ori,jeong2025,calahan2024}, the disks of HD~100546, IRS~48, HD~100453, and tentatively in the MWC~480 disk \citep{Booth2024HD100546, booth2024IRS48, booth2025HD100453, yamato2024}. 
While the estimated \ce{CH3OCHO} excitation temperatures for the V883~Ori and V1057 cyg disks are consistent with thermal desorption ($\sim145$~K and $\sim120$~K, respectively), the rotational temperature of $\sim68$~K we found for the HD~100546 disk is much lower, possibly suggesting a different origin. However, non-LTE conditions inducing sub-thermal excitation could play a role, but a larger number of methyl formate lines and complete information on the collisional rates are needed to robustly infer the excitation conditions.

\subsection{\texorpdfstring{CH$_3$OH/H$_2$O}{CH3OH/H2O} and \texorpdfstring{CH$_3$OCHO/CH$_3$OH}{CH3OCHO/CH3OH} ratios}\label{subsec:ratios}
We leverage the observations of thermally sublimated water vapor at the snowline of the HD~100546 disk \citep{rampinelli2026}, together with multiple detections of methanol lines in the disk, to extract the gas-phase methanol-to-water ratio. We extracted the water column density from the 183~GHz line flux (see Table~\ref{tab:imaging_molecules}), assuming the same integration area and excitation temperature of methanol from the RD analysis ($T_\mathrm{ex} = 100$~K). We obtained a water column density of $(4.6\pm1.2) \times 10^{15}$~\cms and a methanol-to-water ratio of $0.019^{+0.007}_{-0.004}$, with the error bars including uncertainties on the water flux and the methanol column density and excitation temperature. Since water vapor is not detected in the disk of IRS~48 (see Sect.~\ref{subsec:H2O_IRS48}), we extracted a $3\sigma$-lower limit on the methanol-to-water ratio of $\sim0.2$, from the upper limit on the water column density of $\sim1.4\times10^{16}$~\cms, assuming that the emitting area is equal to the beam and an excitation temperature of 103~K as we obtained from the methanol RD analysis. 

Similarly, we extracted the methyl formate-to-methanol ratios in the two disks, leveraging the results of the RD analyses for HD~100546, while we used the band~7 \ce{CH3OH} and \ce{CH3OCHO} column densities calculated at $T = 100$~K by \cite{booth2024IRS48} for IRS~48, since methyl formate is not detected in band~5, thus preventing an RD analysis. The corresponding ratios are $\sim2$ and $0.2$ for the HD~100546 and IRS~48 disks, respectively.
We additionally re-calculated the ratios using the best fit methanol column densities from the non-LTE analysis (see Sect.~\ref{subsec:nonLTE}) and applying the filling factor to the methyl formate and water column densities, founding a consistent result with the ratios calculated from the RD analysis.
The results are presented in Fig.~\ref{fig:CH3OH/H2O} (pink markers refer to the non-LTE results) and compared with other measurements across the star and planet formation timeline. As an additional test, we calculated the \ce{CH3OCHO/CH3OH} ratio in the HD~100546 disk using the \ce{CH3OCHO} column density extracted by assuming the methanol excitation temperature of 100~K. The resulting \ce{CH3OCHO} column density is $9.85\times10^{13}$ and $2.72\times10^{14}$~\cms, when using the band~7 and 5 line fluxes, respectively. The corresponding \ce{CH3OCHO/CH3OH} ratio is $\sim2$ times lower and  $\sim1.5$ higher than the value obtained from the RD analysis, respectively, without changing any interpretation outlined in Sect.~\ref{subsubsec:mf-methanol}.
\section{Discussion}\label{sec:discussion}

\subsection{The COMs-rich but water-poor disk of IRS-48}\label{subsec:H2O_IRS48}
As shown in Sec.~\ref{subsec:methanol}, methanol is detected in the disk of IRS~48 at the location of the mm-crescent both in previous ALMA band~7 \citep{booth2024IRS48,temmink2024} and new band~5 observations presented in this work (see bottom row of Fig.~\ref{fig:M0}). Gas-phase methanol has been proposed to originate from thermal sublimation of methanol ices at the dust crescent \citep{vandermarel2021,booth2024IRS48}. Since water and methanol have similar desorption temperatures \citep{wakelam2017, minissale2022}, we would expect thermal sublimation of water in the same region where thermally desorbed methanol is detected. While water observations in the disk of HD~100546 supports this scenario \citep{rampinelli2026}, the disk of IRS~48 shows only methanol but no water detected at sub-mm wavelengths above the RMS of 5 mJy beam$^{-1}$.

\begin{figure}[ht!]
    \centering
    \includegraphics[scale=0.18]{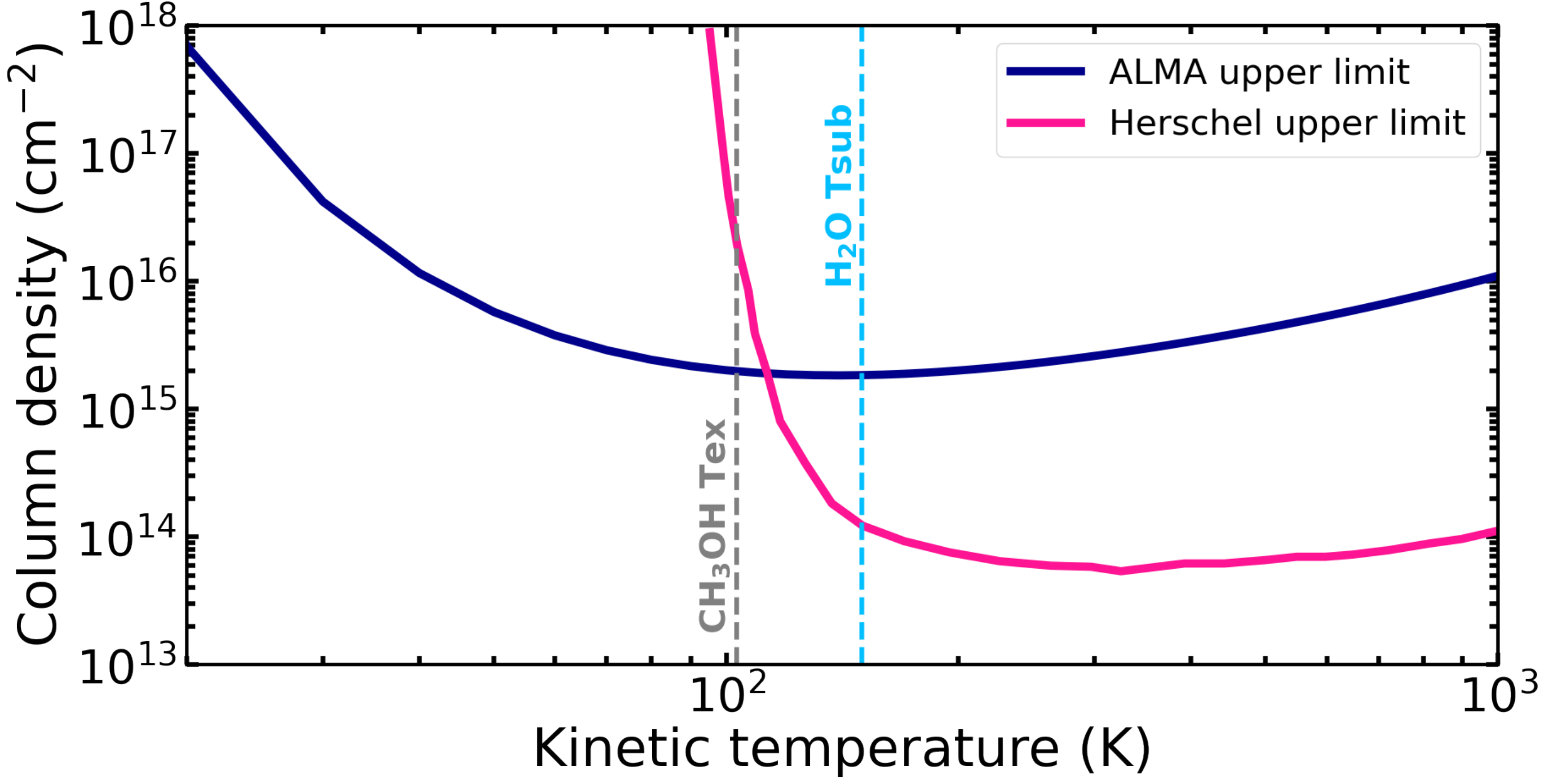}
    \caption{Upper limit on the water column density as a function of kinetic temperature, under the assumption of LTE, from ALMA (blue) and Herschel (pink) non-detections \citep{fedele2013,leemker2023}. Vertical gray and cyan lines indicate the methanol excitation temperature from the RD analysis and the typical water sublimation temperature of $\sim150$~K.}
    \label{fig:upper_water_IRS48}
\end{figure}

Water was previously targeted in the IRS~48 disk by \textit{Herschel} in the DIGIT survey, but it was not detected even at far-infrared wavelengths \citep{fedele2013}. \cite{leemker2023} estimated the maximum water column density above the optically thick dust, under the assumption of local thermodynamic equilibrium, as a function of the assumed kinetic temperature, using the 3$\sigma$ limit from the \textit{Herschel} observations. Figure~\ref{fig:upper_water_IRS48} shows the comparison between the upper limit on the water column density obtained from the non-detection at far-infrared wavelengths (pink line) with the one we similarly calculated from the 3$\sigma$ ALMA upper limit (blue line), assuming that the emitting area is coincident with the mm-crescent ($1.4 \times 10^{-11}$~sr).
If the excitation temperature of water is lower than the one inferred for methanol ($\lesssim 100$~K), the ALMA upper limit (ranging from $10^{15}$ to $10^{18}$~\cms for $T = 20-100$~K) is constraining the water column density. On the other hand, if $T_\mathrm{ex}(\ce{H2O)}$ is higher than $T_\mathrm{ex}(\ce{CH3OH)}$, the upper limit imposed by \textit{Herschel} observations is  constraining the water column density to be lower than $\sim10^{14}$~\cms. In both cases, the low upper limit on the water column density with respect to the high methanol one results in a lower limit on the methanol-to-water ratio which is at least an order of magnitude higher with respect to the one inferred for the disk of HD~100546 (see Sect.~\ref{subsec:methanol} for a detailed discussion), possibly pointing to a distinct ice composition.

However, we note that this discussion is based on the assumption of LTE. 
\cite{rampinelli2026} calculated the critical density $n_\mathrm{crit}$ of the 183~GHz water line as a function of temperature, following \cite{faure2024}. The $n_\mathrm{crit}$ of the water line is between $\sim10^{8}-10^{9}$~\cmt (see also Table~\ref{tab:imaging_molecules} and Appendix~\ref{appendix:critical_density}) while the typical gas density of the IRS~48 disk is expected to be $\lesssim10^8$~\cmt at the mm-crescent \citep{leemker2023}, suggesting that non-LTE effects can also affect water emission.

While ALMA observations have been shown to trace the bulk of the gas-phase water reservoir deeper to the midplane with respect to IR observations \citep{leemker2025HLTau, houge2025}, thanks to the mm-dust being typically more settled than $\mu$m-dust, IRS~48 represents an exception to this picture. The thermochemical model presented by \cite{leemker2023} predicts the $\tau = 1$ layer of the 0.85~mm- and of the 100~$\mu$m-size dust to be at similar heights at the location of the mm-crescent, which limits the potential of ALMA observations to reveal a larger water reservoir with respect to IR wavelengths. 



\subsection{The trail of molecular complexity}\label{subsec:complexity}

\begin{figure*}[ht!]
    \centering
    \includegraphics[scale=0.2]{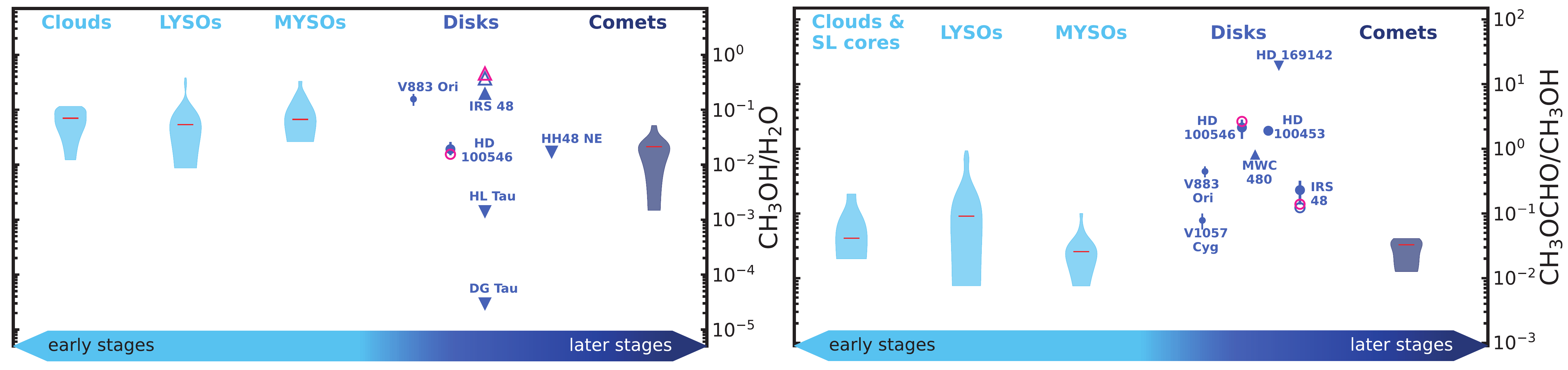}
    \caption{Methanol-to-water (left) and methyl formate-to-methanol (right) ratios across different stages of star and planet formation. The distributions with the corresponding medians (in red) are shown for clouds, starless (SL) cores, LYSOs, MYSOs, and comets from both ice and gas observations. Individual measurements are displayed for disks, including upper and lower limits, from gas-phase observations (see Appendix~\ref{appendix:refs} for the corresponding references). Pink circles refer to ratios extracted after applying the non-LTE analysis to the methanol lines.}
    \label{fig:CH3OH/H2O}
\end{figure*}

\subsubsection{Methanol-to-water ratio}\label{subsubsec:methanol-water}

As shown in the left panel of Fig.~\ref{fig:CH3OH/H2O}, the \ce{CH3OH/H2O} ratio we obtain for HD~100546 is consistent with the lower edge of the distributions of methanol-to-water ratios in clouds, low-mass, and high-mass young stellar objects \citep[LYSOs and MYSOs, see][for an overview, and Appendix~\ref{appendix:refs} for a detailed list of references]{whittet2011,boogert2015,mcclure2023}. This supports the scenario for which the measured methanol-to-water ratio in the disk of HD~100546 is tracing the ice content inherited from early stages of star formation and subsequently released in the gas-phase during the disk phase. However, the methanol-to-water ratio is also consistent with cometary value, which could be indicative of a degree of reprocessing of the methanol and water ice during the disk phase, that can be subsequently reflected in older stages. Nevertheless, we notice that the median of the cometary distribution is within the range of \ce{CH3OH/H2O} ratios measured in younger sources, while the lower cometary ratios correspond to only a small sample of objects, namely 67P/Churyumov-Gerasimenko and a few comets observed within an heliocentric distance of 1~au \citep{bonev2023,biver2023,lippi2024}. Consequently, \citet{lippi2024} emphasizes the need for caution when comparing cometary molecular ratios with those observed in younger objects, as individual measurements or particular cometary classes may not accurately reflect the chemical history of the bulk cometary material, thereby biasing the interpretation.

In the case of the HD~100546 \ce{CH3OH/H2O} ratio, water optical depth could also play a role and multiple water lines are necessary to better constrain the molecular column density and optical depth of water in the disk. If water is optically thicker, its column density is expected to be higher, implying a lower methanol-to-water ratio, less consistent with YSOs, thus supporting the reprocessing scenario. Moreover, while the water vapor emission at 183~GHz is suggested to primarily trace sublimation at the water snowline closer to the midplane with respect to FarIR wavelengths  \citep{pirovano2022,rampinelli2026}, multiple water lines are also needed to definitely assess the thermal desorption origin and rule out a possible contribution from gas-phase formation.

The \ce{CH3OH/H2O} ratio was also extracted for the disk of the outbursting V883 Ori source, from H$_2^{18}$O observations \citep{lee2019,tobin2023,leemker2023}. In this case, the accretion burst drives the midplane condensation front farther out \citep{vanthoff2018,leemker2021,tobin2023,wang2025,yamato2026}, resulting in a methanol-to-water ratio which is also directly tracing the sublimated ice content. In particular, \cite{tobin2023} found a steep drop in the HDO and H$_2^{18}$O column density beyond $\sim80$~au, with water emission extending out to $\sim160$~au, while \cite{yamato2026} observed a steep drop in the methanol emission at $\sim40$~au from band~1 ALMA observations.
In this case, the ratio is consistent with the upper edge of the distributions in early stages. A contribution from a different ice-phase formation route of methanol from radical-radical recombination after water photolysis has been suggested to explain the high methanol-to-water ratio in the V883~Ori source \citep{jeong2025}. This is supported by the high methyl formate-to-methanol ratio in the disk (see the right panel of Fig.~\ref{fig:CH3OH/H2O}), since a large amount of \ce{CH3O}-products are expected in the proposed scenario \citep{shannon2013, ishibashi2021}.

The wide spread of \ce{CH3OH/H2O} values (spanning $\sim$~an order of magnitude) in early stages of star formation may reflect different initial environmental conditions that can be reflected also in the disk stage when the ice content is released in the gas phase. This variety is indeed tentatively reproduced by the two measurements of methanol-to-water ratio currently available in the disk stage, but larger statistics is needed to robustly interpret them, as individual measurements may not be representative of the median of a statistical sample. If we include upper and lower limits, the \ce{CH3OH/H2O} ratios in disks span $\sim5$ order of magnitudes. In particular, the lower limit on the \ce{CH3OH/H2O} ratio in the IRS~48 disk of $\sim0.2$ is the highest value across all objects. As previously discussed in Sect.~\ref{subsec:H2O_IRS48}, IRS~48 is a peculiar water-poor but COMs-rich source. The high methanol-to-water ratio could be indicative of a distinct chemical composition of the ices at the location of the mm-crescent, where efficient vertical mixing can drastically affect the local chemistry \citep{semenov2011}. Moreover, while non-LTE effects at the mm-crescent could play a role \citep{temmink2024}, we do not expect them to be dominant as the non-LTE analysis presented in Sect.~\ref{subsec:nonLTE} shows that the lines are close to LTE.

\cite{soave2026} recently measured an upper limit on the \ce{CH3OH/H2O} ratio in the HL~Tau disk, directly extracted from ALMA observations (see the left panel of Fig.~\ref{fig:CH3OH/H2O}, \citealt{facchini2024}). The low ratio in the HL~Tau disk can be explained by both radiative transfer (depending on the methanol, water, and dust optical depth, \citealt{weaver2018}) and chemical effects (e.g., enhanced water vapor from efficient gas-phase formation at warm temperatures $\gtrsim300$~K that are seen in this disk, \citealt{vandishoeck2014}), or by intrinsically lower methanol abundances \citep{soave2026}. Similarly, the methanol-to-water ratio extracted for the DG~Tau disk from thermochemical modeling of ALMA and \textit{Herschel} observations is $\sim3$ order of magnitudes lower than the median of the cometary distribution \citep[][ and see the left panel of Fig.~\ref{fig:CH3OH/H2O}]{podio2013, podio2019, soave2026}. Such low methanol-to-water ratio could be due to the water vapor emission mainly tracing an abundant water reservoir primarily formed in the gas phase, and/or different wavelength ranges of the observed water and methanol lines resulting in different optical depth effects and/or probed reservoirs \citep{podio2013, soave2026}. Interestingly, the upper limit on the \ce{CH3OH/H2O} ratio in the HH48~NE disk measured from ice observations is consistent with the lower cometary values, possibly linked to some reset in the disk.

This complexity highlights the importance of increasing the number of disks for which the \ce{CH3OH/H2O} ratio can be extracted, in order to interpret individual measurements in light of a larger statistics and effectively compare with early and late stages.


\subsubsection{Methyl formate-to-methanol ratio}\label{subsubsec:mf-methanol}
Similarly to what we did for the methanol-to-water ratio, we extracted the methyl formate-to-methanol ratio in the disk of HD~100546 and compare it to early and late stages of star and planet formation and other available measurements in disks (see the right panel of Fig.~\ref{fig:CH3OH/H2O}, \citealt{mercimek2020,jimenez-serra2025} for a review, and Appendix~\ref{appendix:refs} for more references).

As already highlighted by \cite{booth2024IRS48}, the \ce{CH3OCHO/CH3OH} ratio in disks is typically higher with respect to all the other stages. In particular, while the ratios for the disk of IRS~48 and the disks around the outbursting stars of V883~Ori and V1057~Cyg are within the ranges of YSOs, the disks of HD~100546, HD~100453, HD~169142, MWC~480 show methyl formate-to-methanol ratios higher than earlier and later stages, by even more than an order of magnitude (see Fig.~\ref{fig:CH3OH/H2O} and Appendix~\ref{appendix:refs}). This could be an artifact due to an underestimated methanol column density under the assumption of optically thin emission in disks. However, the non-LTE analysis shows that all the methanol lines are expected to be optically thin in the HD~100546 disk.

We notice that both the \ce{CH3OH/H2O} and \ce{CH3OCHO/CH3OH} ratios obtained from the non-LTE analysis of methanol are consistent with the ones calculated from the RD analysis. This is likely due to the fact that the column density is dominated by the small filling factors ($\sim0.03$ for IR~48 and $\sim0.004$ for HD~100546, see Sect.~\ref{subsec:nonLTE}), which cancels out when computing the ratio, rather than line optical depths. However, we highlight that a robust treatment of line opacities is fundamental, as optically thick lines can lead RD analyses to strongly underestimate molecular column densities. For example, the methanol column density we found for the IRS~48 disk with the non-LTE analysis is a factor of $\sim4$ higher than the RD column density corrected for the filling factor, as a few lines are found to be optically thick. Observations of heavier methanol isotopologues would be particularly useful to observationally address this problem \citep{booth2024IRS48}.


The high methyl formate-to-methanol ratio in disks could also reflect a chemical origin.
In particular, \cite{booth2024IRS48} highlighted the possibility of chemical reprocessing to explain the enhancement of COMs emission in class~II disks. \cite{oberg2009} showed that the reprocessing of methanol ice by UV-photons followed by diffusion and recombination of radicals can support the production of larger COMs such as methyl formate, depending on the branching ratios of \ce{CH3OH} photodissociation, the diffusion barriers of the resulting radicals, and the barriers of reactions producing more complex molecules.
Moreover, \cite{walsh2014} highlighted the impact of photochemistry in determining the abundance of ice- and gas-phase COMs in disks. In particular, the abundance of ice-phase methyl formate in irradiated disks strongly depends on the initial abundances and it is reduced in warm ($\gtrsim30$~K) conditions since the precursor radicals (\ce{CH3O} and \ce{HCO}) are efficiently desorbed. Nonetheless, \cite{walsh2014} showed that gas-phase formation of methyl formate is also effectively sustained by thermal desorption of \ce{CH3OH} and \ce{HCOOH}. In this context, vertical mixing has been shown to strongly affect the abundance of COMs in disks \citep{semenov2011}, which may play a fundamental role in the case of e.g., IRS~48. Moreover, \cite{kakkar2025} showed that the binding energy of methyl formate can be lower than methanol, depending on the underlying crystalline or amorphous water ice model, and on the presence of distinct adsorption sites. This may result in methyl formate efficiently sublimating over a more extended region than methanol, thus affecting column density calculations.
The complexity and variety of physical conditions in protoplanetary disks thus require careful treatment and the interpretation of COMs observations needs the support of further chemical modeling. 

\section{Conclusions}\label{sec:conclusions}
In this work we analyzed ALMA observations of water vapor, methanol, and methyl formate in the chemically rich disks of HD~100546 and IRS~48. In particular, we presented the gas-phase methanol-to-water ratios as a way of revealing the underlying molecular complexity that was built in the ice phase. The main findings can be summarized as follows:
\begin{itemize}
    \item The water vapor line at 183~GHz is not detected in the disk of IRS~48, resulting in a high lower limit on the gas-phase methanol-to-water ratio ($\sim0.2$). The disk of IRS~48 is thus a peculiar COMs-rich but water poor source. 
    \item A rotational diagram analysis of the methanol lines in the HD~100546 disk supports the thermal desorption origin, consistent with the results presented by \cite{evans2025} and what was previously suggested for the water reservoir in the disk \citep{rampinelli2026}. 
    \item The gas-phase methanol-to-water ratio in the disk of HD~100546 ($\sim0.02$) is consistent with the lower edge of the distribution of early stages, possibly indicating an inherited ice reservoir which is subsequently released in the gas-phase in the warmer disk-stage. The methanol-to-water ratio in the disk is also consistent with the cometary distribution. Multiple water lines are needed to robustly assess the water excitation conditions and line opacities.
    \item The non-LTE LVG analysis of the methanol emission in the HD~100546 and IRS~48 disks suggests small emitting areas of $0\farcs12$ and $0\farcs05$ in size, respectively. Methanol lines are found to be close to LTE for both disks. While all methanol lines are consistent with optically thin emission in the disk of HD~100546, a few lines are found to be optically thick for the IRS~48 disk, highlighting the importance of robustly treating line opacities when extracting molecular column densities and excitation conditions.
    \item Methyl formate is detected in band~5 only in the disk of HD~100546. The corresponding methyl formate-to-methanol ratio ($\sim2$) is $\sim$ an order of magnitude higher than younger and older objects. This could reflect a chemical reprocessing of ices supporting the formation of more complex molecules such as \ce{CH3OCHO}. On the other hand, the \ce{CH3OCHO/CH3OH} ratio in the disk of IRS~48 ($\sim0.2$) is consistent with the distribution in younger objects.
\end{itemize}

These findings underscore the potential of ALMA line emission observations of water vapor and complex organic molecules to reveal the molecular complexity stored in ice reservoirs in the outer regions of planet-hosting disks. Moreover, they offer new insights into the chemical history of molecular complexity by bridging the gap between early and late stages of star and planet formation. The results presented in this work pave the way for a systematic characterization of the basic ingredients of molecular complexity, water and COMs, in the birthplaces of planets.

\begin{acknowledgments}
This paper makes use of the following ALMA data: 

ADS/JAO.ALMA 2023.1.01431.S.

ADS/JAO.ALMA 2021.1.00738.S.

ALMA is a partnership of ESO (representing its member states), NSF (USA) and NINS (Japan), together with NRC (Canada), MOST and ASIAA (Taiwan), and KASI (Republic of Korea), in cooperation with the Republic of Chile. The Joint ALMA Observatory is operated by ESO, AUI/NRAO and NAOJ.

L.R., S.F., and M.L. are funded by the European Union (ERC, UNVEIL, 101076613). Views and opinions expressed are however those of the authors only and do not necessarily reflect those of the European Union or the European Research Council. Neither the European Union nor the granting authority can be held responsible for them. S.F. also acknowledges financial contribution from PRIN-MUR 2022YP5ACE.
A.S.B. is supported by a Clay Postdoctoral Fellowship from the Smithsonian Astrophysical Observatory.
P.C. acknowledges support by the ANID BASAL project FB210003.
M.B. has received funding from the European Research Council (ERC) under the European Union’s Horizon 2020 research and innovation programme (PROTOPLANETS, grant agreement No. 101002188).
L.T. acknowledges financial support from the European Research Council via the ERC Synergy Grant “ECOGAL” (project ID 855130).

\end{acknowledgments}





%
\facilities{ALMA}

\software{\texttt{CASA} v6.5.4 \citep{CASA2022},  
          \texttt{numpy} \citep{harris2020}, 
          \texttt{matplotlib} \citep{hunter2007}, \texttt{bettermoments} \citep{teague2019bettermoments}, \texttt{gofish} \citep{teague2019gofish}
          }


\appendix
\renewcommand{\thefigure}{A.\arabic{figure}}
\setcounter{figure}{0}
\renewcommand{\thetable}{A.\arabic{table}}
\setcounter{table}{0}

\section{Non detection of water vapor in IRS 48}\label{appendix:water_IRS48}
\begin{figure*}[ht!]
    \centering
    \includegraphics[scale=0.2]{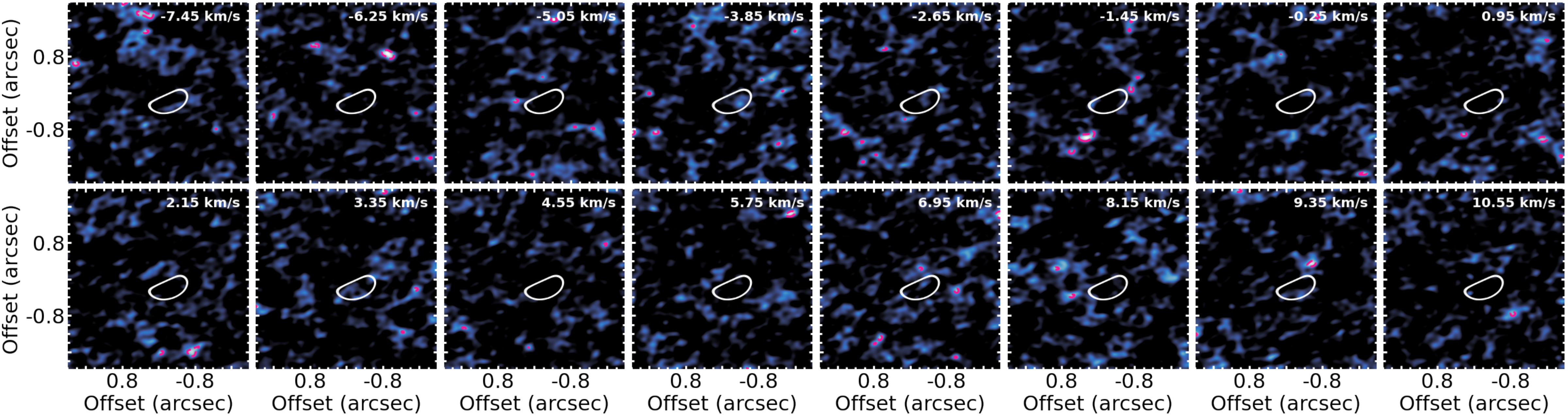}
    \caption{Channel maps of the non-detected water line at 183~GHz. The systemic velocity is 4.55~\kms. Pink contours show 3$\sigma$ levels. The cube was imaged with natural weighting and a channel width of 1.2~\kms.}
    \label{fig:channel_maps}
\end{figure*}

Figure~\ref{fig:channel_maps} shows the channel maps of the non-detected 183~GHz water line in the IRS~48 disk. We imaged the cube with natural weighting to maximize the sensitivity. We tested different channel width (0.6 and 1.2~\kms), and a $uv$ taper of $0\farcs2$ to increase the sensitivity to extended emission. In all the tested cases water is non detected.

\begin{figure*}[ht!]
    \centering
    \includegraphics[scale=0.18]{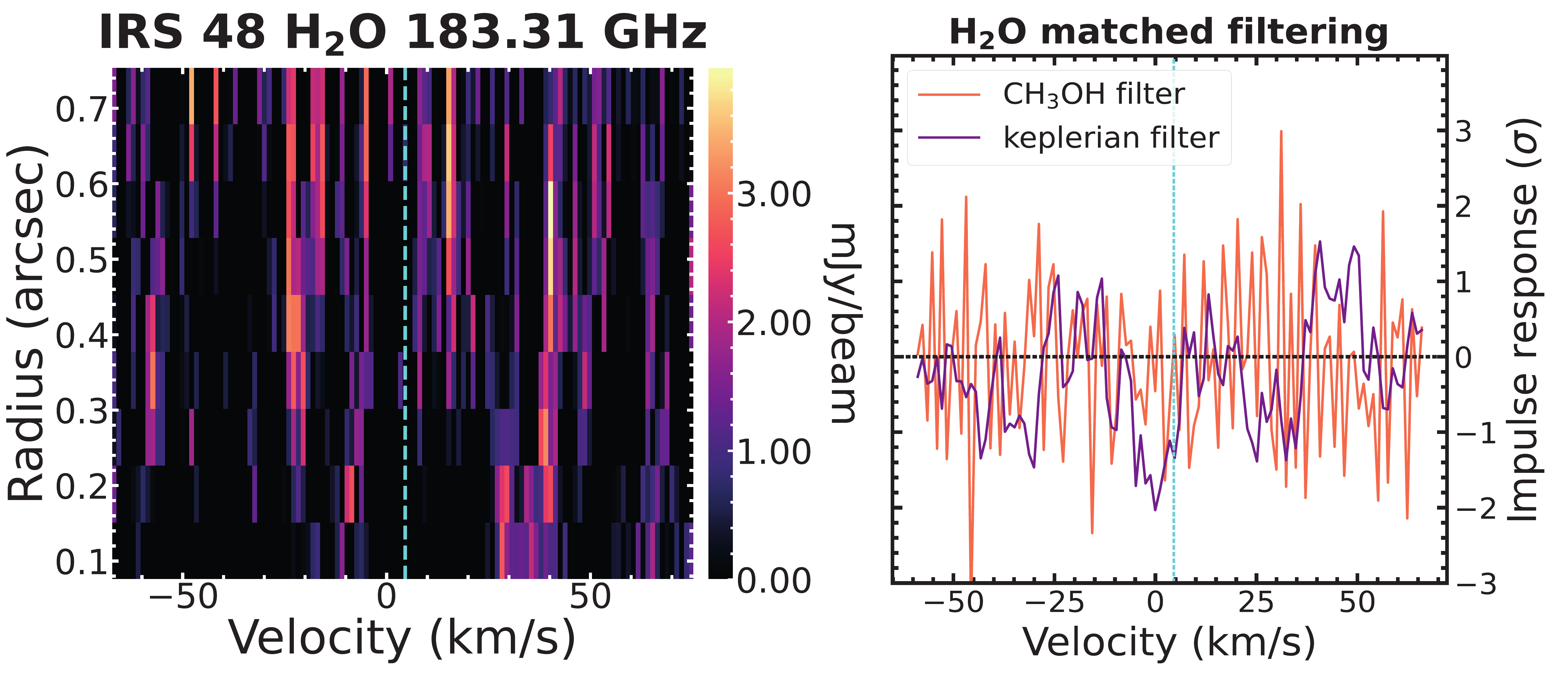}
    \caption{Left: water line intensity (non detection) as a function of velocity (x-axis) and radius (y-axis). Right: Filter response to the water spectrum using matched filtering in the $uv$-plane. the dashed cyan lines indicate the systemic velocity of 4.55~\kms.}
    \label{fig:teardrop_visible_H2O_IRS48}
\end{figure*}

The left panel of Fig.~\ref{fig:teardrop_visible_H2O_IRS48} shows the water line intensity as a function of velocity (x-axis) and radius (y-axis), extracted from a mask centered at the location of the mm-crescent and after correcting for the Keplerian rotation. The emission pattern from a Keplerian disk shows the typical \textit{teardrop} shape as in the bottom row of Fig.~\ref{fig:teardrop_methanol}. 
The right panel of Fig.~\ref{fig:teardrop_visible_H2O_IRS48} shows the attempt to detect the water line using the matched filter analysis implemented in the \texttt{VISIBLE} package \citep{loomis2018visible}, which consists of spanning the sampled visibility with a model filter. We tested two filters: one directly using the imaged cube of the detected methanol line at 173.54~GHz (orange line) and a keplerian model (purple line). Both filters responses do not match any water emission at the systemic velocity of 4.55~\kms (dashed light blue line).

\section{Teardrop plots of the methanol lines}\label{appendix:teardrop_methanol}

\begin{figure*}[ht!]
    \centering
    \includegraphics[scale=0.1]{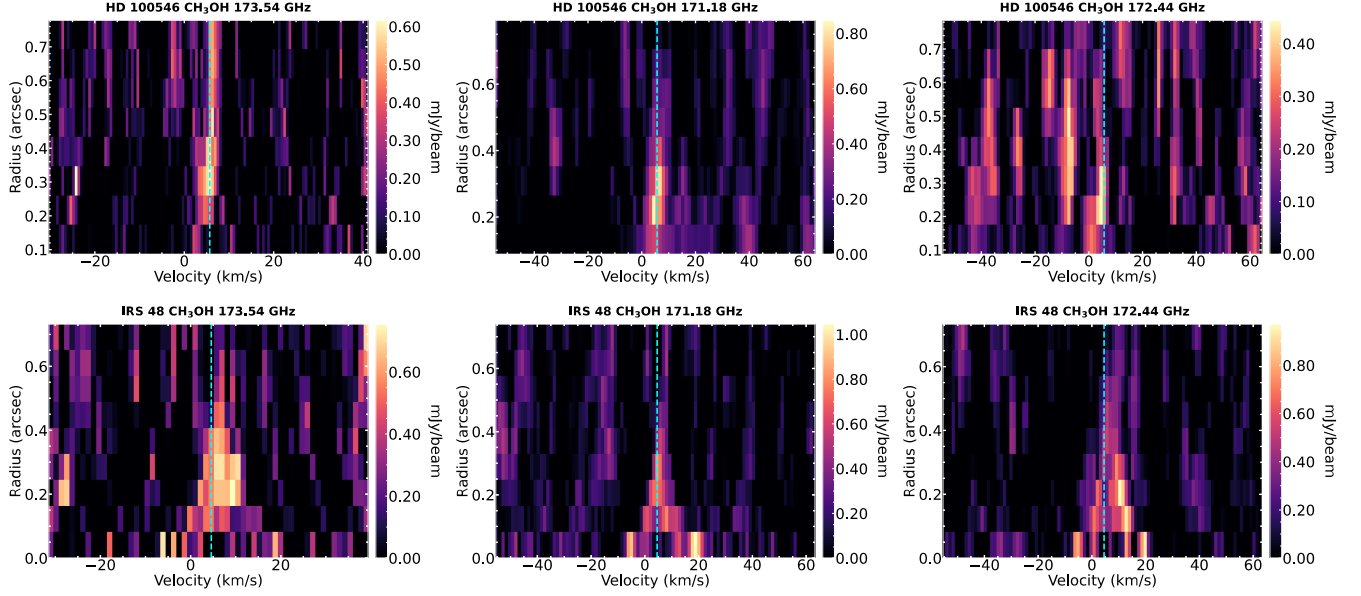}
    \caption{Line intensity as a function of velocity (x-axis) and radius (y-axis) of the three methanol lines, for the two sources analyzed in this work. The vertical light blue line indicates the systemic velocity of 5.7~\kms and 4.55~\kms for HD~100546 and IRS~48, respectively.}
    \label{fig:teardrop_methanol}
\end{figure*}

In Fig.~\ref{fig:teardrop_methanol} we show the so-called \textit{teardop plots} of the methanol lines for HD~100546 (top row) and IRS~48 (bottom row) disks. 
The resulting plots for HD~100546 show the typical emission pattern expected for a disk in Keplerian rotation, with a larger linewidth at smaller radii and the emission centered at the systemic velocity of 5.7~\kms (dashed light blue line). The top right panel shows the non detection of the methanol line at 172.44~GHz in the HD~100546 disk. For the disk around IRS~48 all three lines are detected, with the emission pattern being slightly de-centered from the systemic velocity of 4.55~\kms (see also the bottom left panel of Fig.~\ref{fig:spectrum_rotdiag_CH3OH}), since the plot was extracted from a spatial mask centered on the mm-crescent where the peak of the methanol emission is (see bottom panels of Fig.~\ref{fig:M0}), rather than centered on the star location.

\section{Band 7 methanol and methyl formate lines}\label{appendix:b7_methanol}
Table~\ref{tab:b7_methanol} shows an overview of the methyl formate and methanol lines detected in previous ALMA band~7 observations \citep{Booth2024HD100546,booth2024IRS48, temmink2024,evans2025} for the IRS~48 disk and by \cite{evans2025} for the HD~100546 disk. We list the line parameters (upper state energy, degeneracy, and Einstein coefficients) from the CDMS catalog, and the line fluxes.

\begin{table*}
  \caption{Methanol and methyl formate lines from previous band~7 observations used in this work \citep{booth2024IRS48,temmink2024,evans2025}.}
  \label{tab:b7_methanol}
  \resizebox{\textwidth}{!}{%
  \begin{threeparttable}
    \noindent
          \begin{tabular}{c c c c c c c}
              \hline
              \hline
              Transition\tnote{a} & Frequency & $E_\mathrm{u}$ & log(A$_\mathrm{ul}$) & $g_\mathrm{u}$ & Line flux\tnote{b} & $n_\mathrm{cr}$\\
              & [GHz] & [K] & [s$^{-1}$] & & [mJy~km~s$^{-1}$] & [\cmt]\\
              \hline
              & & HD 100546 & & & \\
              \hline
              \ce{CH3OH} ($7_{0,7}-6_{0,6}$ E) & 338.1245 & 78.08 & -3.771 & 60 & $105 \pm 19$ & $1.7\times10^6$\\
              \ce{CH3OH} ($7_{1,7}-6_{1,6}$ E) & 338.3446 & 70.55 & -3.778 & 60 & $96 \pm 19$ & $2\times10^6$\\
              \ce{CH3OH} ($7_{0,7}-6_{0,6}$ A) & 338.4087 & 64.98 & -3.769 & 60 & $137 \pm 19$ & $2\times10^6$ \\
              \ce{CH3OH} ($7_{3,4}-6_{3,3}$ E) & 338.5600 & 127.71 & -3.854 & 60 & $51 \pm 19$ & $1\times10^7$\\
              \ce{CH3OH} ($7_{3,5}-6_{3,4}$ E) & 338.5832 & 112.71 & -3.856 & 60 & $103 \pm 19$ & $6\times10^6$\\
              \ce{CH3OH} ($7_{1,6}-6_{1,5}$ E) & 338.6149 & 86.05 & -3.767 & 60 & $110 \pm 19$ & $8\times10^5$\\
              \ce{CH3OH} ($7_{2,5}-6_{2,4}$ A) & 338.6398 & 102.72 & -3.802 & 60 & $69 \pm 19$ & $4\times10^6$\\
              \ce{CH3OH} ($14_{1,13}-14_{0,14}$ A) & 349.1070 & 260.20 & -3.356 & 116 & $75 \pm 19$ & $3\times10^6$\\
              \ce{CH3OH} ($1_{1,1}-0_{0,0}$ A) & 350.9051 & 16.84 & -3.480 & 12 & $45 \pm 19$ & $4\times10^5$\\
              \ce{CH3OH} ($13_{0,13}-12_{1,12}$ A) & 355.6029 & 211.03 & -3.597 & 108 & $51 \pm 19$ & $4\times10^6$\tnote{d}\\
              \ce{CH3OCHO} ($32_ {0,32}-31_{0,31}$ A) & 344.0296 & 276.08 & -3.208 & 130 & & -\\
              \ce{CH3OCHO} ($32_ {1,32}-31_{1,31}$ A) & 344.0296 & 276.08 & -3.208 & 130 & $63\pm10$\tnote{e} & - \\
              \ce{CH3OCHO} ($32_ {0,32}-31_{0,31}$ E) & 344.0293 & 276.10 & -3.192 & 130 & & -\\
              \ce{CH3OCHO} ($32_ {1,32}-31_{1,31}$ E) & 344.0293 & 276.10 & -3.192 & 130 & & -\\
              \hline
              & & IRS 48 & & & \\
              \hline
              \ce{CH3OH} ($3_{3,1}-4_{2,3}$ E) & 337.1359 & 61.64 & -4.801 & 28 & $20.0 \pm 3.4$ & $4\times10^6$ \\
              \ce{CH3OH} ($7_{0,7}-6_{0,6}$ A) & 338.1245 & 78.08 & -3.771 & 60 & $67.5 \pm 4.1$ & $7\times10^5$\\
              \ce{CH3OH} ($7_{6,1}-6_{6,0}$ E) & 338.4310 & 253.95 & -4.343 & 60 & $9.7 \pm 2.5$ & $1\times10^8$\tnote{d}\\
              \ce{CH3OH} ($7_{6,1}-6_{6,0}$ A) & 338.4424 & 258.70 & -4.344 & 60 & $14.3 \pm 3.2$ & $5\times10^7$\tnote{d}\\
              \ce{CH3OH} ($7_{6,2}-6_{6,1}$ A) & 338.4424 & 258.70 & -4.344 & 60 & $14.3 \pm 3.2$& $5\times10^7$\tnote{d}\\
              \ce{CH3OH} ($7_{5,3}-6_{5,2}$ E) & 338.4565 & 189.00 & -4.078 & 60 & $22.6 \pm 3.2$& $2\times10^7$\tnote{d}\\
              \ce{CH3OH} ($7_{5,2}-6_{5,1}$ E) & 338.4752 & 201.06 & -4.078 & 60 & $33.4 \pm 3.4$& $3\times10^7$\tnote{d}\\
              \ce{CH3OH} ($7_{5,3}-6_{5,2}$ A) & 338.4863 & 202.89 & -4.076 & 60 & $18.6 \pm 3.1$& $5\times10^7$\tnote{d}\\
              \ce{CH3OH} ($7_{5,2}-6_{5,1}$ A) & 338.4863 & 202.89 & -4.076 & 60 & $18.6 \pm 3.1$& $5\times10^7$\tnote{d}\\
              \ce{CH3OH} ($7_{4,4}-6_{4,3}$ E) & 338.5041 & 152.89 & -3.940 & 60 & $42.1 \pm 3.6$& $7\times10^6$\\
              \ce{CH3OH} ($7_{4,3}-6_{4,2}$ E) & 338.5303 & 160.99 & -3.938 & 60 & $43.8 \pm 3.8$& $3\times10^7$\\
              \ce{CH3OH} ($7_{3,4}-6_{3,3}$ E) & 338.5600 & 127.71 & -3.854 & 60 & $48.4 \pm 3.6$& $1\times10^7$\\
              \ce{CH3OH} ($7_{3,5}-6_{3,4}$ E) & 338.5832 & 112.71 & -3.856 & 60 & $56.9 \pm 3.8$& $6\times10^6$\\
              \ce{CH3OH} ($7_{2,5}-6_{2,4}$ A) & 338.6398 & 102.72 & -3.802 & 60 & $57.3 \pm 3.4$& $4\times10^6$\\
              \ce{CH3OH} ($2_{2,0}-3_{1,3}$ A) & 340.1411 & 44.67 & -4.556 & 20 & $13.2 \pm 2.8$& $2\times10^6$\\
              \ce{CH3OH} ($13_{1,12}-13_{0,13}$ A) & 342.7298 & 227.47 & -3.374 & 108 & $64.0 \pm 3.0$& $3\times10^6$\\
              \ce{CH3OH} ($14_{1,13}-14_{0,14}$ A) & 349.1070 & 260.20 & -3.356 & 116 & $58.7 \pm 3.0$& $3\times10^6$\\
              \ce{CH3OH} ($1_{1,1}-0_{0,0}$ A) & 350.9051 & 16.84 & -3.480 & 12 & $51.2 \pm 3.5$& $4\times10^5$\\
              \ce{CH3OH} ($9_{5,4}-10_{4,6}$ E) & 351.2365 & 240.51 & -4.438 & 76 & $25.3 \pm 2.8$& $3\times10^7$\tnote{d}\\
              \ce{CH3OH} ($13_{0,13}-12_{1,12}$ A) & 355.6029 & 211.03 & -3.597 & 108 & $61.9 \pm 3.3$& $2\times10^6$\tnote{d}\\
              \ce{CH3OH} ($15_{1,14}-15_{0,15}$ A) & 356.0072 & 295.26 & -3.337 & 124 & $49.6 \pm 3.0$& $4\times10^6$\tnote{d}\\
              \hline
              \hline
          \end{tabular}
            \begin{tablenotes}\centering
            \footnotesize
              \item[a] See Table~\ref{tab:imaging_molecules} for quantum numbers formatting.
              \item[b] Peak intensity at the brightest pixel for IRS~48 \citep{temmink2024}, disk integrated flux within 110~au for HD~100546 \citep{evans2025}.
              \item[c] Critical density calculated for a kinetic temperature of $100$~K. The critical density for the methyl formate lines is not calculated because collisional rates are unavailable.
              \item[d] Critical density calculated for collisions with para-\ce{H2} only, as collisional rates with ortho-\ce{H2} are not available.
              \item[e] The reported flux corresponds to the sum of the blended methyl formate lines.
          \end{tablenotes}
  \end{threeparttable}
  }
\end{table*} 

\section{Critical densities}\label{appendix:critical_density}
\begin{figure*}[ht!]
    \centering
    \includegraphics[scale=0.6]{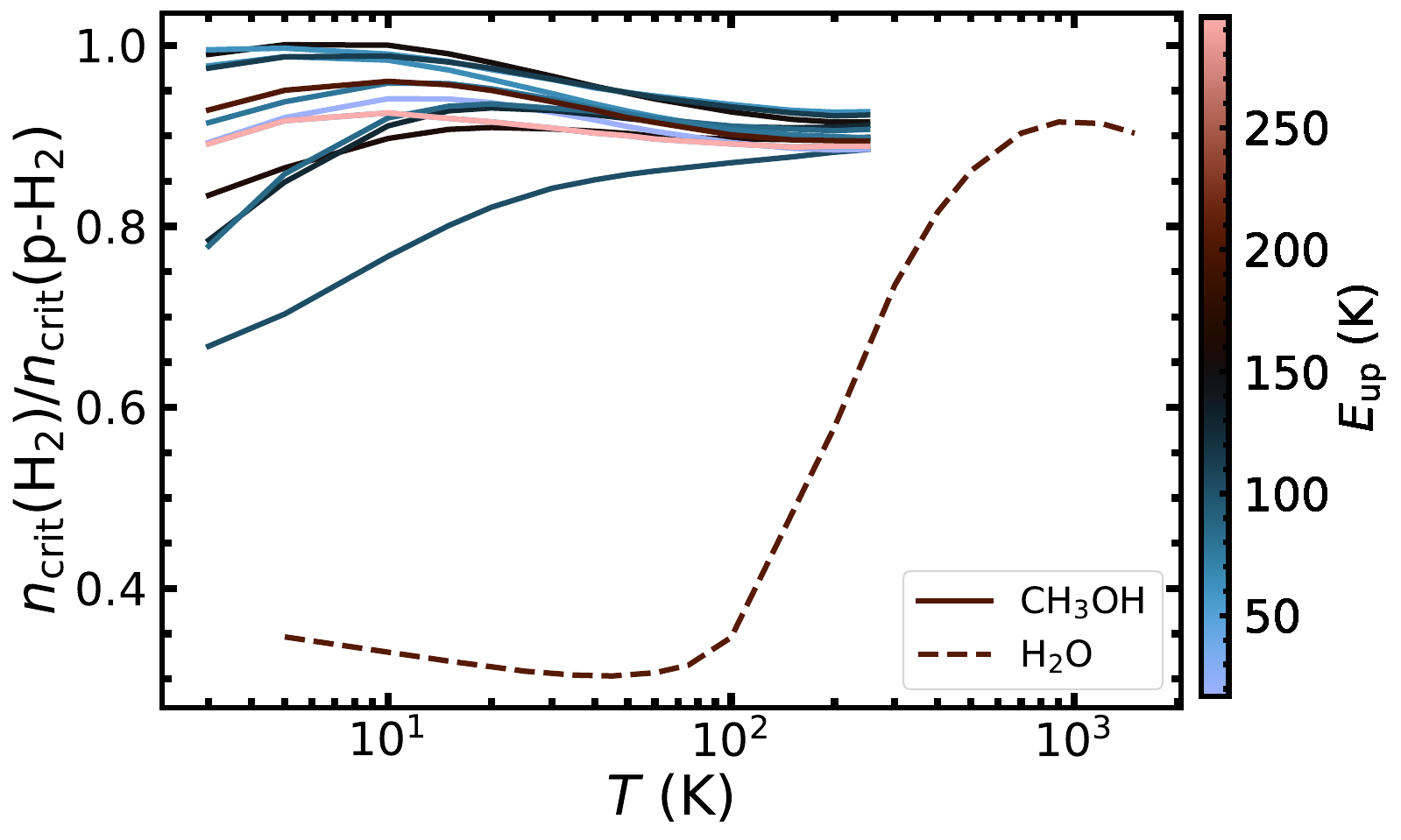}
    \caption{Ratio between the \ce{H2} and para-\ce{H2} critical densities as a function of temperature, for the methanol (solid) and water (dashed) lines listed in Table~\ref{tab:imaging_molecules} and \ref{tab:b7_methanol}. The colorbar shows the upper state energy of each transition.}
    \label{fig:ncr}
\end{figure*}

We calculated the critical densities of the methanol and water lines listed in Table~\ref{tab:imaging_molecules} and \ref{tab:b7_methanol} as described in Appendix~C of \cite{rampinelli2026}, following \cite{faure2024}. Collisional rates with para-\ce{H2} are taken from \cite{rabli_rotational_2010} and are calculated for temperatures between $10$ and $200$~K, while those with both ortho- and para-\ce{H2} are taken from \cite{dagdigian2024} and are calculated for $3<T<250$~K. The collisional rates of water with both para- and ortho-\ce{H2} are calculated by \cite{daniel2010,daniel2011}. Line lists and related collisional rates can be found on the LAMDA, BASECOL, or EMAA databases \citep{lamda2005,dubernet_basecol2012:_2013,faure2025}. We provide the total \ce{H2} critical density for those lines with available collisional rates with both para- and ortho-\ce{H2}, assuming an ortho-to-para ratio of 3. When only collisional rates with para-\ce{H2} are provided in the literature, we calculate the critical density of para-\ce{H2} only. Figure~\ref{fig:ncr} shows the ratio between \ce{H2} and para-\ce{H2} critical densities as a function of temperature for methanol and water transitions used in this paper. While para-\ce{H2} is the dominant collider for methanol, ortho-\ce{H2} dominates the critical density for the water transition at 183~GHz up to $\sim100$~K, with a difference that is within a factor of $\sim3$.

\section{Methanol and methyl formate across star and planet formation} \label{appendix:refs}

A list of references from which the methanol-to-water and methyl formate-to-methanol ratios presented in this work were obtained is provided. However, this list is not exhaustive, as many of the cited works report measurements that originate from earlier studies. We thus encourage the reader to consult the reference lists within those papers for the original sources. For broader context and comprehensive overviews of the topic, the reader may also refer to the following reviews: \cite{boogert2015, mercimek2020, jimenez-serra2025}.

Literature references:
\begin{itemize}
    \item \textbf{Low-mass young stellar objects (YSOs):} \cite{bottinelli2010, whittet2011, jorgensen2016, bergner2017}; \citeyear{bergner2019}; \cite{oberg2018, jacobsen2019, belloche2020, manigand2020, bianchi2020, yang2021, rocha2024, chen2024}.
    \item \textbf{High-mass young stellar objects (MYSOs):} \cite{whittet2011, coletta2020, chen2023, li2024}.
    \item \textbf{Class~II disks and FU~Ori sources:} \cite{lee2019, tobin2023, leemker2023, calahan2024, yamato2024, jeong2025, booth2025HD100453, booth2026}.
    \item \textbf{Comets:} \cite{oberg2011ice, mumma2011, leroy2015, bockelee-morvan2000, boogert2015, rubin2019, lippi2024}
\end{itemize}

\bibliography{sample7}{}
\bibliographystyle{aasjournalv7}



\end{document}